\documentclass[utf8,]{FrontiersinHarvard} 
\usepackage{url,lineno}
\usepackage[onehalfspacing]{setspace}
\newcommand{\dm}[1]{}
\newcommand{\bb}[1]{}
\newcommand{\gk}[1]{}
\newcommand{\ita}{\textit{a}} 
\newcommand{\scatterheight}{0.22\textheight}
\newcommand{\mapwidth}{0.48\linewidth}

\def\keyFont{\fontsize{8}{11}\helveticabold }
\def\firstAuthorLast{Moffat {et~al.}} 
\def\Authors{
    David Moffat$^{1,2,\dagger*}$, 
    Angus Laurenson$^{1,2,\dagger}$,
    Victor Martinez-Vicente$^{1}$,
    Gemma Kulk$^{1,2}$,
    Xuerong Sun$^{3}$, 
    Robert J. W. Brewin$^{3}$, 
    and Shubha Sathyendranath$^{1,2}$
}

\def\Address{$^{1}$Plymouth Marine Laboratory, Plymouth, UK \\
$^{2}$National Center for Earth Observation, Plymouth Marine Laboratory, UK \\
$^{3}$Department of Earth and Environmental Sciences, Centre for Geography and Environmental Science, University of Exeter, Cornwall, United Kingdom}
\def\corrAuthor{David Moffat}
\def\corrEmail{dmof@pml.ac.uk}
\begin{document}
\onecolumn
\firstpage{1}

\title[Beyond chlorophyll: machine learning estimates of diagnostic pigments]{Beyond chlorophyll: machine learning estimates of diagnostic phytoplankton pigments from multispectral ocean colour data} 

\author[\firstAuthorLast ]{\Authors} 
\address{} 
\correspondance{} 
 \extraAuth{}

\maketitle
\begingroup \renewcommand{\thefootnote}{\fnsymbol{footnote}} \footnotetext[2]{These authors contributed equally to this work.} \endgroup

\begin{abstract}
Phytoplankton play a central role in marine ecosystems and the global carbon cycle, with their functions in ocean biogeochemical cycles differing types of phytoplankton. Though standard techniques exist for monitoring the concentration of phytoplankton from ocean colour data, their community composition remains difficult to observe at large scales. The main phytoplankton pigment chlorophyll-\textit{a}, readily available from satellite-based ocean-colour data, is widely used as a measure of phytoplankton biomass but provides limited information on taxonomic composition, in and by itself. Accessory pigments, some of which are considered to be diagnostic of some important phytoplankton groups, offer additional information on phytoplankton communities, but their retrieval from ocean-colour remote-sensing data is challenging due to the limited spectral bands traditionally available from satellite data and due to their strong covariance with chlorophyll-\textit{a}.
In this study, we evaluate the capability of machine learning methods to estimate diagnostic pigment concentrations from multispectral satellite observations. Using a global dataset of 33,640 High Performance Liquid Chromatography (HPLC) measurements matched with ESA Ocean Colour Climate Change Initiative (OC-CCI) ocean-colour reflectance data, we compare Random Forest and TabPFN models trained on multispectral reflectance with baseline models using chlorophyll-\textit{a} alone. A temporally stratified validation scheme is employed to reduce the effects of autocorrelation.
Results show that multispectral models consistently outperform approaches using only satellite derived chlorophyll, demonstrating that ocean colour reflectance contains additional information relevant to pigment discrimination. Improvements vary by pigment, with pigments strongly correlating with chlorophyll\textit{-a} showing limited gains, while others exhibit substantial improvement. These findings highlight the potential of machine learning to extract ecologically relevant information from satellite data beyond conventional chlorophyll-based methods.
{\tiny \keyFont{ \section{Keywords:} machine learning, diagnostic pigment, phytoplankton, ocean colour, remote sensing, phytoplankton community composition, explainable AI } 
}
\end{abstract}

\section*{Preprint} 
\date{20 August 2026}\\
This manuscript is a non-peer-reviewed preprint submitted to ArXiv. The manuscript has also been submitted for peer review to \textit{Frontiers in Marine Science}. The content of this preprint has not yet undergone formal peer review and should therefore be considered preliminary. Subsequent versions of this manuscript may differ from the version presented here following peer review and publication.

Citation: Moffat, D. et al. (2026). Beyond chlorophyll: machine learning estimates of diagnostic phytoplankton pigments from multispectral ocean colour data. ArXiv.

\section{Introduction}
Phytoplankton are a diverse group of microscopic organisms that live in the sunlit portions of aquatic ecosystems, both freshwater and marine. Through photosynthesis, they are responsible for around 50\% of global primary production of organic carbon~\citep{longhurst1995estimate,field-1998}. They form the basis of the food web upon which marine life depends~\citep{falkowski2012ocean}. Understanding how phytoplankton respond to climate change and what effects this will have on the environment are important questions that require long-term monitoring, on a global scale~\citep{sathyendranath2017ocean}. Ocean-colour satellite observations provide an important tool for such monitoring, based on the observation of the pigment chlorophyll-\textit{a}~\citep{gordon1980phytoplankton,occci}. All phytoplankton contain chlorophyll-\textit{a}, representing the sum of monovinyl chlorophyll-\textit{a}, divinyl chlorophyll-\textit{a}, and chlorophyllide-\textit{a}. Chlorophyll-\textit{a} is essential for photosynthesis. In the open-ocean, Case-1 waters, it is generally assumed that chlorophyll-\textit{a} concentration can be treated as the single, independent variable that determines the reflectance signal (all other substances that contribute to the signal are assumed to covary in some fashion with chlorophyll-\textit{a}); and chlorophyll-\textit{a} can be accurately measured from remote sensing. It is a measure of phytoplankton biomass and informs models of marine primary production ~\citep{platt-1988,behrenfeld-1997,kulk-2020,baxter2025water,silsbe2025global,sathyendranath2020reconciling}. The Case-1 assumption that chlorophyll-\textit{a} concentration can be an indicator of the underlying ecosystem structure has been exploited to infer community structure from chlorophyll-\textit{a}~\citep{sun2025coupling, brewin2015influence} with remarkable success. However, we recognise that deviation from the Case-1 assumption can occur even in the open ocean. 
Furthermore, estimates of phytoplankton biomass, productivity, and community structure become more uncertain towards the coast where other constituents in the water can dominate the optical signal.

Quantifying the contributions from individual phytoplankton types to total phytoplankton biomass is not a trivial problem: Phytoplankton are a highly diverse community, with estimates suggesting between 5,000 and 150,000 species worldwide~\citep{sournia1991marine,armbrust2015uncovering} spanning multiple phyla~\citep{tett-1995}. Several methods exist for identifying individual phytoplankton species \textit{in situ}, including DNA sequencing and microscopy~\citep{BracherBoumanScientificRoadmap2017,ioccg-2014}. However, these techniques are expensive and typically have limited geographical coverage. Simplifications are therefore necessary and phytoplankton are often grouped by their function in the ecosystem into Phytoplankton Functional Types (PFTs; ~\citealp{nair2008remote,ioccg-2014}) and by their physical size into Phytoplankton Size Classes (PSCs)~\citep{brewin2014multicomponent}. A method that is commonly used to identify different groups of phytoplankton \textit{in situ}, is High Performance Liquid Chromatography (HPLC). This method measures phytoplankton pigments that can then be used to estimate PFTs with established methods such as Diagnostic Pigment Analysis (DPA; \citealp{vidussi-2001}), ChemTax~\citep{mackey-1996}, and more recently PhytoClass~\citep{hayward_span_2023}. Yet, the relationship between pigments and phytoplankton types is ambiguous, as several phytoplankton groups can synthesise the same pigments, and phytoplankton are able to modify the relative concentrations of their pigment complement according to physiological state or environmental conditions (\citealp{nair2008remote}; Table \ref{tab:pigments}). Accessory pigments help phytoplankton adapt to underwater light conditions by altering the spectrum of light they can absorb for photosynthesis~\citep{majchrowski2000influence,dubinsky2010light}. The varied morphology and ecological niches of phytoplankton groups have led to the evolution of different accessory pigments~\citep{reynolds2006ecology}. These broadly correspond to PFTs; for example, the pigment peridinin is associated with dinoflagellates, whilst \textit{Prochlorococcus} uniquely uses divinyl-chlorophylls (Table \ref{tab:pigments}). But, the pigment fucoxanthin, often associated with diatoms, can also be synthesised by other phytoplankton groups (Table \ref{tab:pigments}). It has been demonstrated that four broad groups of phytoplankton can be discriminated from diagnostic pigments at the global scale: cyanobacteria, diatoms/dinoflagellates, haptophytes, and green algae~\citep{kramer_how_2019}. 

\begin{table}[ht]
\small
    \caption{\small Overview of pigments estimated by the machine learning models and their taxonomic affiliation with phytoplankton groups. Information taken from~\citet{nair2008remote}, with the taxonomic affiliation identified for those phytoplankton groups in which the specific pigment is always or often present. Some pigments are unique markers to certain phytoplankton groups (i.e., unambiguous) and others appear across multiple phytoplankton groups (i.e., ambiguous).}
    \vspace{0.3cm}
    \begin{tabular}{l|l|>{\raggedright\arraybackslash}p{9.7cm}}
        \textbf{Pigment} & \textbf{Marker type} & \textbf{Taxonomic affiliation} \\
        \hline \hline 
        \vspace{1pt}
        19'-butanoyloxyfucoxanthin & Ambiguous & Haptophytes, Pelagophytes \\
        19'-hexanoyloxyfucoxanthin & Ambiguous & Chrysophytes, Haptophytes \\
        Alloxanthin & Unambiguous & Cryptophytes \\
        Monovinyl Chlorophyll-\ita & Ambiguous & All phytoplankton, except \textit{Prochlorococcus} \\
        Divinyl chlorophyll-\ita & Unambiguous & \textit{Prochlorococcus} \\
        Chlorophyll-b & Ambiguous & Chlorophytes, Euglenophytes, Prasinophytes \\
        Divinyl chlorophyll-\textit{b} & Ambiguous & \textit{Prochlorococcus} \\
        Fucoxanthin & Ambiguous & Chrysophytes, Diatoms, Haptophytes, Pelagophytes, Raphidophytes \\
        Peridinin & Unambiguous & Dinoflagellates \\
        Zeaxanthin & Ambiguous & Cyanobacteria, Chlorophytes, Chrysophytes, Euglenophytes, Prasinophytes, \textit{Prochlorococcus}, Raphidophytes \\
    \end{tabular}
    \label{tab:pigments}
\end{table}

Several studies have explored machine learning approaches for predicting phytoplankton diversity indicators, such as pigments, phytoplankton types, and size classes, from satellite observations and environmental data. Early work by \citet{brewin_intercomparison_2011} compared nine approaches for estimating PSCs and found that abundance-based methods, which rely solely on chlorophyll-\textit{a}, performed comparably to approaches incorporating spectral or environmental variables such as sea surface temperature. This result raises questions about the extent to which additional predictors provide independent information beyond chlorophyll-\textit{a}. Subsequent studies have attempted to leverage spectral information more explicitly. \citet{bracher_using_2015} developed an empirical model based on linear regression of truncated principal components derived from hyperspectral reflectance, trained using coincident HPLC pigment measurements. When applied to multispectral observations of the Medium Resolution Imaging Spectrometer (MERIS), the model achieved good performance. Similarly, \citet{el2019estimation} applied self-organising maps to predict pigments at the global scale and reported strong performance. However, the use of random data splitting for validation in these studies does not account for spatial and temporal autocorrelation, and may therefore lead to overly optimistic estimates of predictive skill, as the training and validation data may not be statistically independent. \citet{stock_accuracy_2020} conducted a systematic comparison of machine learning approaches for predicting diagnostic pigment concentrations normalised by total chlorophyll-\textit{a}. They demonstrated that the choice of cross-validation strategy has a strong influence on estimated model performance, emphasising the importance of accounting for the spatial and temporal structure of oceanographic data. More recently, Foundation Models were developed for remote sensing products, but the models were validated only for the prediction of chlorophyll-\textit{a}~\citep{dawson2025sentinel}.

In this study, we developed machine learning models to predict phytoplankton pigments from ocean colour observations of the Ocean Colour Climate Change Initiative (OC-CCI;~\citealp{occci}). A key unresolved question is whether multispectral ocean colour observations contain information on phytoplankton pigment composition beyond that provided by chlorophyll-\textit{a}. Because many accessory pigments are strongly correlated with chlorophyll-\textit{a}, apparent predictive skill may arise simply from covariance with phytoplankton biomass rather than from independent spectral information. Our hypothesis is that multispectral reflectance observations contain information that can discriminate accessory pigments beyond their covariance with chlorophyll-\textit{a}. To test this hypothesis, we train two machine learning models using remote sensing reflectances ($R_{rs}$) at six OC-CCI wavebands, together with $K_{d}(490)$ and chlorophyll-\textit{a}. As a control, we also train a baseline model using remote sensing derived chlorophyll-\textit{a} as the sole predictor. This baseline represents abundance based approaches and provides a benchmark against which the additional information contained in multispectral observations can be evaluated. We validate all models using temporally independent \textit{in situ} observations selected to minimise correlation with the training dataset and apply the resulting models at the global scale to investigate whether ecologically meaningful patterns in phytoplankton pigment composition can be recovered from long-term ocean colour data records.

\section{Materials \& Methods}
\subsection{\textit{In situ} data}
\textit{In situ} data are required to train and evaluate model performance. We use a dataset of 33,640 High Performance Liquid Chromatography (HPLC) measurements, collated from multiple sources between 1991 and 2021. Full details of the \textit{in situ} dataset are provided in \cite{sun2026coupling}. In the purposes of the rest of the paper we refer to total chlorophyll-\ita (Tchla), i.e. both monovinyl and divinyl. Quality assurance was applied to the \textit{in situ} data by removing values below 0.005 mg m$^{-3}$ where data were only reported to three decimal places. This filtering was applied where measurements approached the sensor detection limit, in order to remove quantisation artefacts.

\begin{figure}[b]
    \centering
    \includegraphics[width=1.05\linewidth]{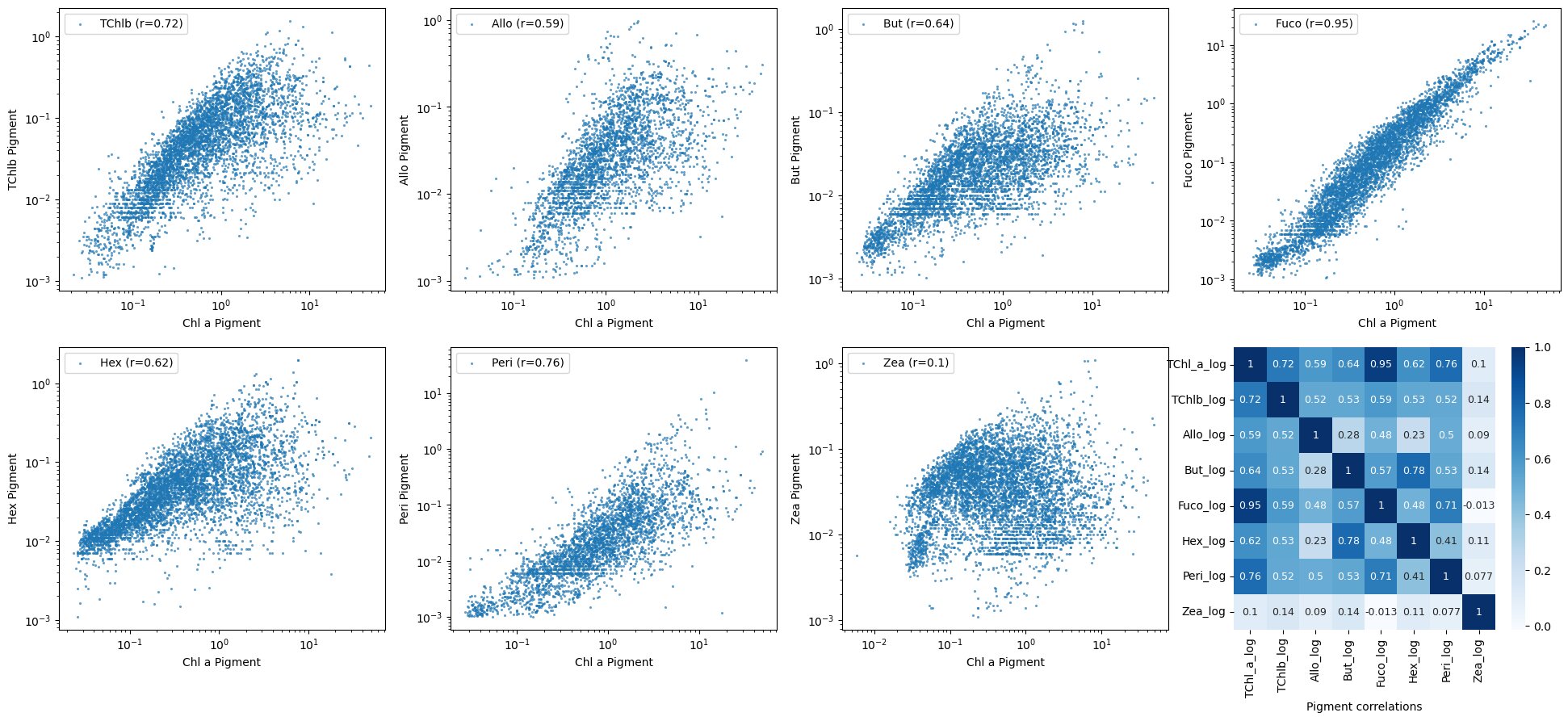}
    \caption{\small Relationships between chlorophyll\textit{-a} and other pigments in the \textit{in situ} dataset, including the Pearson correlation in log space. The bottom right figure shows the Pearson correlation in log space of each pigment compared with each other pigment.}
    \label{fig:pig_corr}
\end{figure}

It is generally expected that the proportions of accessory pigments are highly correlated and not truly independent~\citep{bricaud2004natural}. A correlation plot of the pigments in the \textit{in situ} dataset is shown in Figure~\ref{fig:pig_corr}. It is shown that most pigments in the \textit{in situ } dataset exhibited a strong correlation with chlorophyll-\textit{a}. The highest correlation was observed for fucoxanthin, which was nearly perfectly correlated with chlorophyll-\textit{a}, with a Pearson correlation coefficient of $r = 0.95$. Zeaxanthin was the only exception, showing little correlation with chlorophyll-\textit{a} ($r = 0.10$).

\subsection{Earth Observation data}

The HPLC dataset spans several decades, exceeding the lifespan of individual ocean colour satellite missions. Therefore, data from multiple sensors must be merged to form a consistent dataset. Here, we use the European Space Agency (ESA) Ocean Colour Climate Change Initiative (OC-CCI) version 6.0 dataset at 4-km spatial and daily temporal resolution ~\citep{occci} for match-up with \textit{in situ} HPLC observations. OC-CCI data were also used for application of the developed pigment models at the global scale, for which a 4-km, monthly image of May 2010 was used, as an month, with 2010 representing a year with good data coverage in OC-CCI, which includes multiple sensors contributing to the merged data product.

HPLC measurements were matched with coincident remote sensing reflectances (\textit{$R_{rs}$}), the attenuation coefficient of light at 490 nm ($K_{d}(490)$) and chlorophyll\textit{-a} concentrations from OC-CCI to create a combined \textit{in situ} and EO match-up dataset. Observations were matched when they occurred on the same day and the \textit{in situ} location fell within the satellite pixel. Only \textit{in situ} samples collected within the upper 5 m of the water column were considered. Where multiple samples existed at the same location and day but at different depths within the top 5 m, pigment concentrations were averaged. Quality control was applied to the match-up dataset using a 3$\times$3 pixel window centred on each \textit{in situ} observation. Match-ups were retained only when at least four of the nine satellite pixels contained valid observations and the coefficient of variation within the window was less than 0.15, thereby reducing the influence of sub-pixel spatial heterogeneity and residual retrieval errors~\citep{jordan2023spatial}.

\begin{figure}[h]
    \centering
    \includegraphics[]{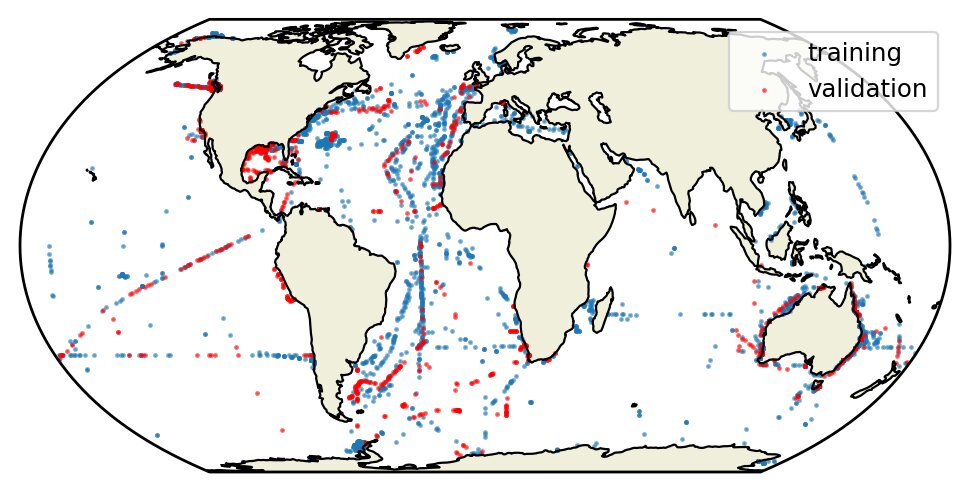}
    \caption{\small Global map of match-up dataset. Blue points show locations of training data, whilst red points show data held back for validation, from years 2002, 2007, 2012, and 2017}
    \label{fig:global_map}
\end{figure}

\subsection{Data Partitioning}
\textit{In situ} measurements are often spatially and temporally correlated. Research cruises collect sequences of observations along transects, resulting in strong correlations due to geographic proximity and seasonal sampling. Furthermore, sampling is often biased towards particular regions (e.g., the Atlantic Ocean) and seasons (e.g., summer). To account for this, we partitioned the match-up dataset into training and validation subsets by withholding four years of data (2002, 2007, 2012, and 2017) for validation. The remaining 20 years were used for training. This resulted in a total of 6,625 data points available for training and 1,701 data points for validation. We note that the number of match-up data points available for each individual pigment varies from these total numbers.

Partitioning the data by year enables assessment of model generalisation to unseen temporal conditions and partially mitigates the effects of temporal autocorrelation. However, it may introduce distributional differences between training and validation datasets, particularly due to geographic variability in cruise sampling. Figure~\ref{fig:global_map} shows the spatial distribution of the match-up dataset and its partitioning between training and validation sets. The training data cover several regions not included in the validation set, such as the vicinity of Madagascar and the western Pacific Ocean. Conversely, some regions are present only in the validation dataset, such as the Gulf of Mexico, as such we need to consider how this method will extrapolate to these locations.

\subsection{Machine learning}
\label{machine-learning}
We compare two machine learning approaches for predicting pigment concentrations from spectral information. The first approach is a foundation model for tabular data, TabPFN~\citep{hollmann2023tabpfn}. TabPFN is a transformer-based foundation model trained on large numbers of synthetic tabular datasets and uses Bayesian posterior estimation to make predictions without explicit model retraining. TabPFN has been shown to perform well on small tabular datasets and avoids the need for extensive model-specific training. The second approach is a Random Forest (RF) model~\citep{breiman2001random}. RF models can capture non-linear relationships and are relatively robust to overfitting and noise, making them suitable for small and noisy datasets~\citep{opitz1999popular}. RF models have also performed well in related ocean colour applications~\citep{toming2026towards,laine2024machine}. To assess the contribution of spectral information (i.e., \textit{Rrs}, $K_{d}(490)$), we also trained a RF baseline model using only chlorophyll-\textit{a} as input. This provides a benchmark for comparison with abundance-based approaches, which relate the PFT or PSC to chlorophyll-\textit{a} reported in literature.

All models were trained to predict the concentrations of eight pigments: chlorophyll-\textit{a}, chlorophyll-\textit{b}, alloxanthin, 19'-butanoyloxyfucoxanthin, 19'-hexanoyloxyfucoxanthin, fucoxanthin, peridinin, and zeaxanthin (Table \ref{tab:pigments}). All target variables were log-transformed to better represent both low and high concentration ranges. The predictor variables used for both TabPFN and RF models included $R_{rs}$($\lambda$) at 412, 443, 490, 510, 560 and 665 nm, attenuation coefficient at 490 nm ($K_{d}(490)$) and chlorophyll-\textit{a} (i.e., all from OC-CCI). Informal testing was performed using different subsets of predictor features, eg., Spectral $R_{rs}$ only, and it was decided to include ($K_{d}(490)$) and chlorophyll-\textit{a} as this provided improvements to the results. Note that the RF baseline model was trained using chlorophyll-\textit{a} only.

Model hyperparameters were optimised separately for each pigment. For TabPFN, the variant with automated hyperparameter optimisation (HPO) was used, with 50 optimisation iterations per pigment model. For the Random Forest models, hyperparameters including the number of trees, number of features, number of training samples, and maximum tree depth were optimised using a grid search.

Model performance was evaluated using the coefficient of determination ($R^2$; ~\citealt{ozer1985correlation}). The $R^2$ statistic quantifies the proportion of variance in the observed pigment concentrations that is explained by the model predictions, with a value of 1 indicating perfect agreement, a value of 0 indicating performance equivalent to predicting the mean of the observations, and negative values indicating performance worse than the mean predictor. As pigment concentrations were modelled in log-transformed space, all reported $R^2$ values were calculated using log-transformed observations and predictions.

\begin{table}[t]
    \caption{\small Overview of the different machine learning models developed to predict eight phytoplankton pigments using ocean colour satellite observations. Details of the models are provided in Section~\ref{machine-learning}. The maximum total number of observations (\textit{N}) used for training and validation of the models is also provided, but specific numbers vary per pigment, due to not all pigments being captured for all samples. }
    \vspace{0.3cm}
    \resizebox{0.95\textwidth}{!}{
    \begin{tabular}{l|llcc}
        \textbf{Model} & \textbf{Method} & \textbf{Predictor variables} & \textbf{Max \textit{N} (Training)} & \textbf{Max \textit{N} (Validation)} \\
        \hline \hline 
        \vspace{1pt}
        TabPFN & Tabular Foundation Model & $R_{rs}$, $K_{d}(490)$, Chl-\textit{a} & 6,625 & 1,701 \\
        RF & Random Forest model & $R_{rs}$, $K_{d}(490)$, Chl-\textit{a} & 6,625 & 1,701 \\
        Baseline & Random Forest model & Chl-\textit{a} & 6,625 & 1,701 \\
    \end{tabular}}
    
    \label{tab:ml_methods}
\end{table}

\subsection{Explainable AI analysis}

To interpret the machine learning models, we employed SHapley Additive exPlanations (SHAP)~\citep{lundberg2017unified,lundberg2020local}, a model-agnostic method based on cooperative game theory that quantifies the contribution of each input feature to individual model predictions. SHAP values represent the marginal contribution of each feature relative to a baseline prediction, enabling consistent and locally accurate attribution of model outputs. SHAP analysis was applied to the trained RF models for each pigment to assess the relative importance of predictor variables, including $R_{rs}$($\lambda$), chlorophyll-\textit{a}, and $K_{d}(490)$. SHAP violin plots were used to visualise the distribution of feature contributions across the dataset, showing both the magnitude and direction of influence on pigment predictions. Such plots allow identification of which spectral bands contribute most strongly to each pigment, and whether their effects are consistent across observations. This provides insight into the extent to which the models rely on chlorophyll-\textit{a} versus independent spectral information, and helps to characterise the nature of the pigment signals captured by the machine learning models.

\section{Results}

\subsection{Comparison of pigment models performance}

\begin{figure}[tbh]
\centering

\includegraphics[width=0.85\linewidth,height=\scatterheight,keepaspectratio]{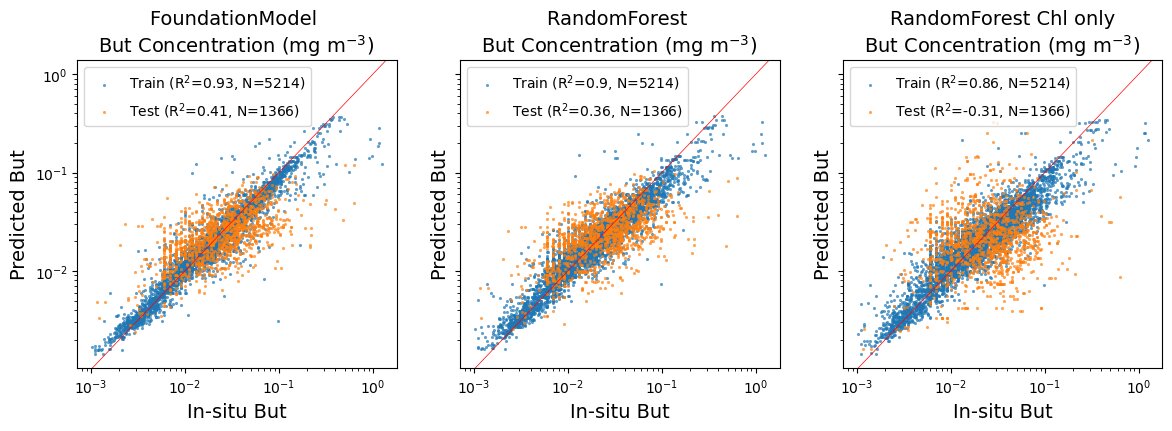}\\
\small (a) 19'-butanoyloxyfucoxanthin\\
\vspace{0.4em}
\includegraphics[width=0.85\linewidth,height=\scatterheight,keepaspectratio]{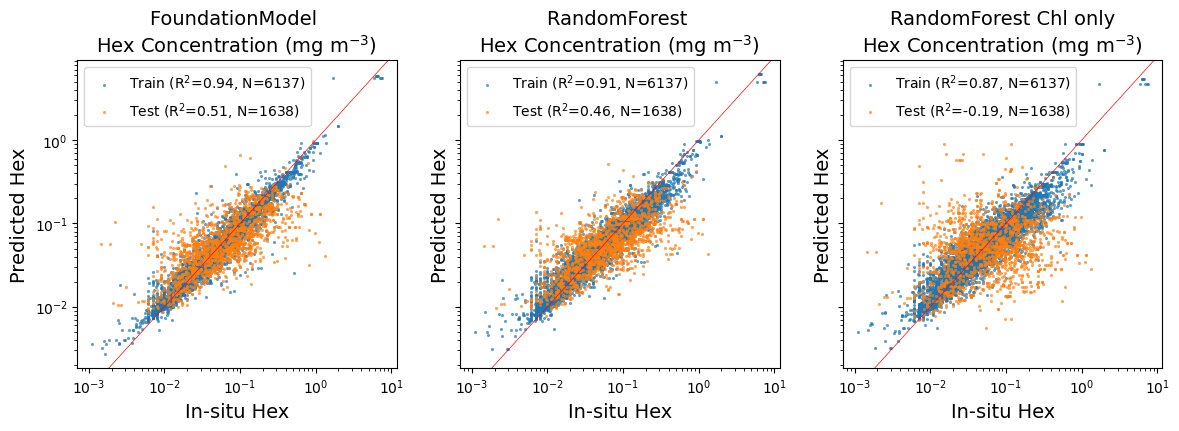}\\
\small (b) 19'-hexanoyloxyfucoxanthin \\
\vspace{0.4em}
\includegraphics[width=0.85\linewidth,height=\scatterheight,keepaspectratio]{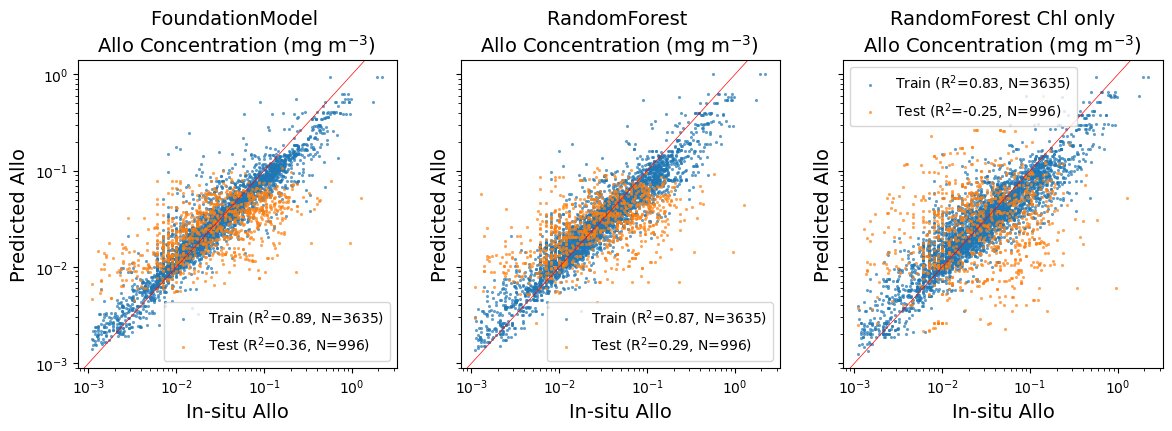}\\
\small (c) Alloxanthin \\
\vspace{0.4em}
\caption{\small Scatter plots of predicted versus measured pigment concentrations for 19'-butanoyloxyfucoxanthin (a), 19'-butanoyloxyfucoxanthin (b), alloxanthin (c). The x-axis shows HPLC-derived measurements and the y-axis shows model predictions of each pigment in mg m$^{-3}$. Results for the training (blue) and validation (orange) datasets are presented for each type of model, i.e., the TabPFN (a,d,g) and RF (b,e,h) models that incorporate spectral information and the baseline model (c,f,i) that is based on chlorophyll-a as a predictor.}
\label{fig:result_scatter_1}
\end{figure}

\begin{figure}[hp]
\centering
\includegraphics[width=0.85\linewidth,height=\scatterheight,keepaspectratio]{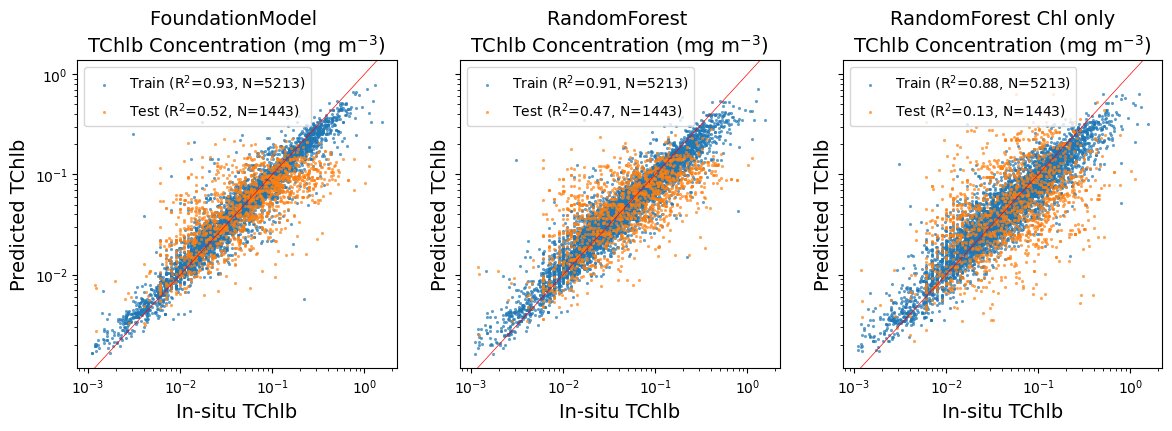}\\
\small (a) Chlorophyll\textit{-b} \\
\vspace{0.4em}
\includegraphics[width=0.85\linewidth,height=\scatterheight,keepaspectratio]{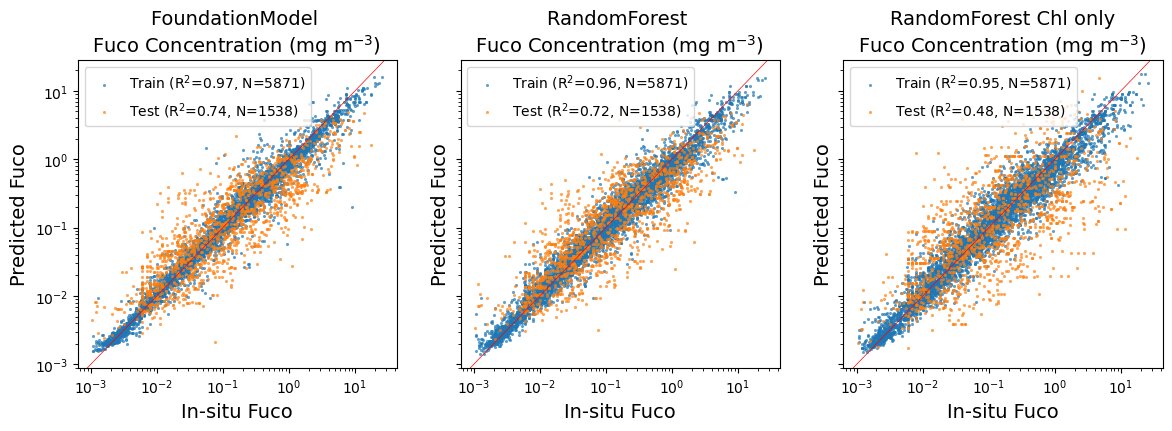}\\
\small (b) Fucoxanthin \\
\vspace{0.4em}
\includegraphics[width=0.85\linewidth,height=\scatterheight,keepaspectratio]{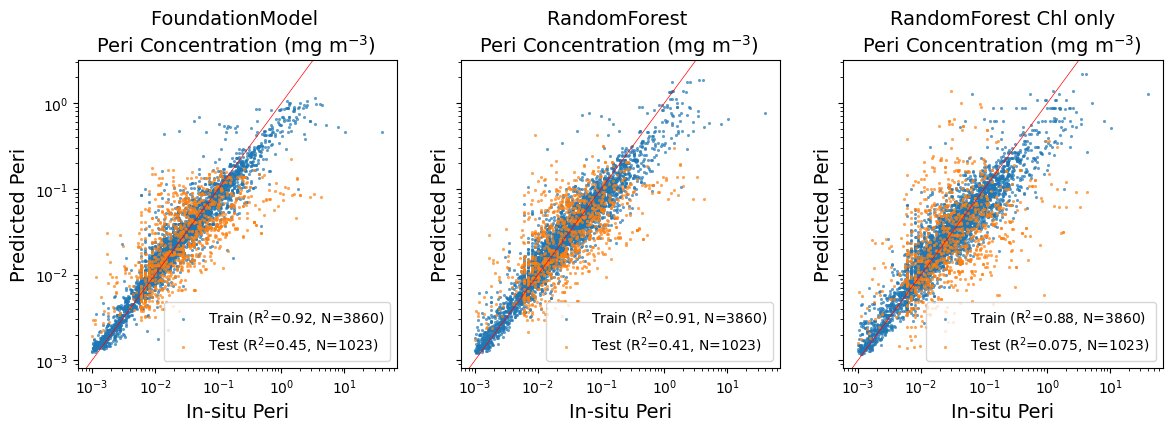}\\
\small (c) Peridinin \\
\vspace{0.4em}
\includegraphics[width=0.85\linewidth,height=\scatterheight,keepaspectratio]{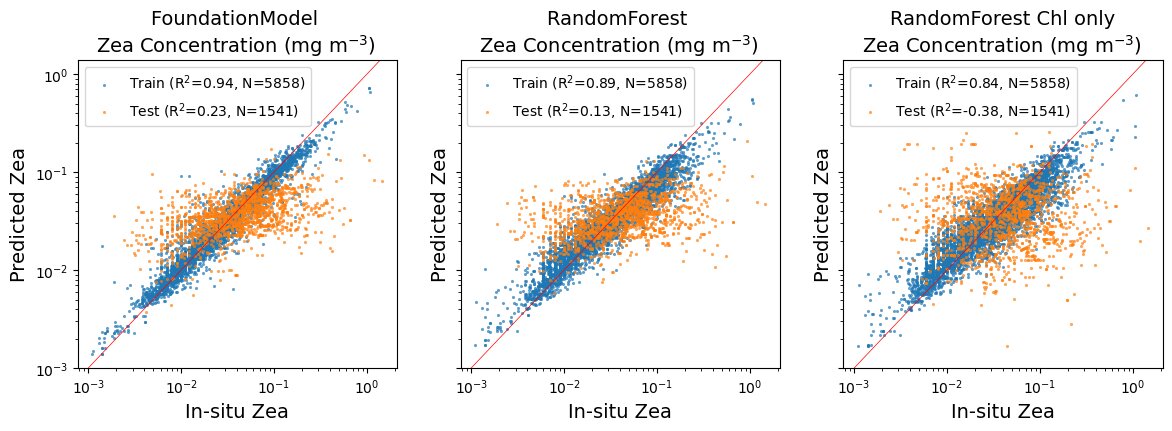}\\
\small (d) Zeaxanthin \\
\caption{\small As in Figure~\ref{fig:result_scatter_1}, for chlorophyll\textit{-b} (a), fucoxanthin (b), peridinin (c), and zeaxanthin (d).}
\label{fig:result_scatter_2}
\end{figure}

\begin{figure}[htbp]
\centering
\begin{tabular}{p{0.49\textwidth} p{0.49\textwidth}}
\includegraphics[width=0.48\textwidth]{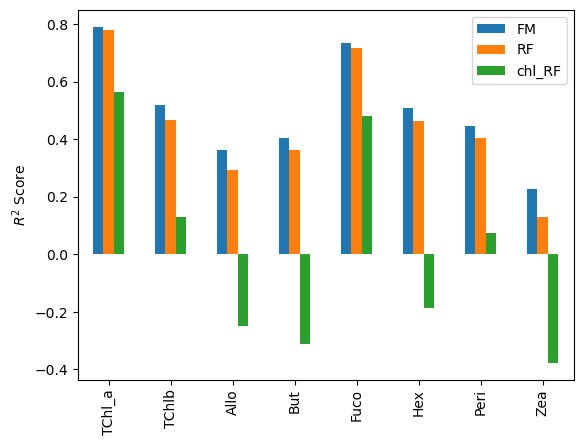} &
        \includegraphics[width=0.47\textwidth]{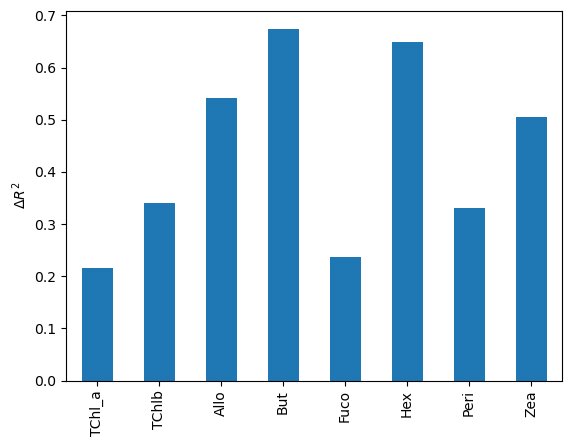} \\ 
\centering \small  (a) Validation $R^2$ scores  & 
\centering \small (b) Improvement relative to chlorophyll-only RF model \\

\end{tabular}
\caption{\small Comparison of model performance on the validation dataset. Panel (a) shows the validation $R^2$ values obtained for each pigment and model configuration. Panel (b) shows the improvement in $R^2$ achieved by the Random Forest model using multispectral reflectance relative to the chlorophyll-only Random Forest baseline, illustrating the additional predictive information contained within the reflectance observations.}
\label{fig:r2_comparison}
\end{figure}

The performance of each pigment prediction model is summarised in Figures~\ref{fig:result_scatter_1} and ~\ref{fig:result_scatter_2}, with results shown for the training and validation datasets for the TabPFN foundation model and the RF model using spectral information and the baseline model using chlorophyll-a as input only. All models showed relatively high performance during training, with $R^2$ ranging between 0.89 -- 0.97 for the TabPFN model, 0.87 -- 0.95 for the RF model and 0.83 -- 0.95 for the baseline model across the eight different pigments. When pigment prediction was compared with independent validation data, performance was generally lower compared to that of the training dataset with $R^2$ ranging between 0.23 -- 0.74 for the TabPFN model, 0.13 -- 0.72 for the RF model and 0.075 -- 0.48 for the baseline model. The TabPFN model for each pigment showed highest performance, closely followed by the RF model that also incorporated spectral information (Figure~\ref{fig:r2_comparison}a), and these two models also showed similar bias and root-mean-square-difference (Table~\ref{tab:full_metrics}). The models for chlorophyll-a and fucoxanthin showed highest performance and those for zeaxanthin showed lowest performance across all model types (Figure~\ref{fig:r2_comparison}a) Further metrics were calculated, beyond coefficient of determination, however there was very little difference between the different models using these metrics, as shown in Table~\ref{tab:full_metrics}. 

\begin{table}
    \caption{\small Comparison of phytoplankton pigment retrieval performance for the foundation model (predFM), chlorophyll-only random forest (predRF chl), and full random forest (predRF). Statistics shown are sample size (n), Pearson correlation (r), bias, root mean square difference (RMSD), Centre-pattern root mean square difference (CP-RMSD), and coefficient of determination (R$^2$). }
    \vspace{0.1cm}
    \centering

    \resizebox{0.65\textwidth}{!}{
    \begin{tabular}{ll|ccccccccc}
    Pigment & Model & n & r & bias & RMSD & CP-RMSD & R$^2$ \\
    \hline \hline
    TChlb & predFM & 6656 & 0.76 & 3.22 & 3.41 & 1.14 & 0.52 \\
    TChlb & predRF chl & 6656 & 0.71 & 3.21 & 3.41 & 1.15 & 0.13 \\
    TChlb & predRF & 6656 & 0.76 & 3.21 & 3.41 & 1.15 & 0.47 \\
    Allo & predFM & 4631 & 0.65 & 3.68 & 3.86 & 1.18 & 0.36 \\
    Allo & predRF chl & 4631 & 0.58 & 3.68 & 3.86 & 1.18 & -0.25 \\
    Allo & predRF & 4631 & 0.62 & 3.68 & 3.86 & 1.18 & 0.29 \\
    But & predFM & 6580 & 0.73 & 4.00 & 4.12 & 0.97 & 0.41 \\
    But & predRF chl & 6580 & 0.62 & 4.00 & 4.12 & 0.98 & -0.31 \\
    But & predRF & 6580 & 0.70 & 4.00 & 4.12 & 0.97 & 0.36 \\
    Fuco & predFM & 7409 & 0.60 & 2.79 & 3.22 & 1.61 & 0.74 \\
    Fuco & predRF chl & 7409 & 0.57 & 2.76 & 3.22 & 1.66 & 0.48 \\
    Fuco & predRF & 7409 & 0.59 & 2.77 & 3.22 & 1.63 & 0.72 \\
    Hex & predFM & 7775 & 0.45 & 3.12 & 3.28 & 0.99 & 0.51 \\
    Hex & predRF chl & 7775 & 0.35 & 3.12 & 3.28 & 1.01 & -0.19 \\
    Hex & predRF & 7775 & 0.41 & 3.12 & 3.27 & 1.00 & 0.46 \\
    Peri & predFM & 4883 & 0.65 & 3.76 & 4.01 & 1.41 & 0.45 \\
    Peri & predRF chl & 4883 & 0.56 & 3.76 & 4.02 & 1.42 & 0.08 \\
    Peri & predRF & 4883 & 0.61 & 3.76 & 4.01 & 1.41 & 0.41 \\
    Zea & predFM & 7399 & 0.73 & 3.45 & 3.58 & 0.94 & 0.23 \\
    Zea & predRF chl & 7399 & 0.64 & 3.45 & 3.58 & 0.95 & -0.38 \\
    Zea & predRF & 7399 & 0.71 & 3.45 & 3.58 & 0.95 & 0.13 \\
    \end{tabular}}
    \label{tab:full_metrics}
\end{table}

The baseline model, using chlorophyll-\textit{a} as input, consistently underperformed relative to those models incorporating multispectral reflectance (i.e., TabPFN and RF models; Figure~\ref{fig:r2_comparison}a). As shown previously in Figure~\ref{fig:pig_corr}, several pigments exhibited strong correlations with chlorophyll-\textit{a}. Therefore, the improvement observed when including reflectance bands can indicate the extent to which additional spectral information contributes beyond these correlations, which is shown in Figure~\ref{fig:r2_comparison}b. While the models for fucoxanthin showed high performance ($R^2$ $>$ 0.72), the TabPFN and RF models exhibited the smallest improvement over the baseline model for this pigment. This is consistent with Figure~\ref{fig:pig_corr}, which shows that fucoxanthin was highly correlated with chlorophyll-\textit{a} and we might therefore expect that this pigment can largely be predicted from abundance alone. The 19'-butanoyloxyfucoxanthin and 19'-hexanoyloxyfucoxanthin pigment models showed intermediate performance ($R^2$ between 0.36 -- 0.51), but exhibited the largest improvement relative to the baseline model (Figure~\ref{fig:r2_comparison}b). This was primarily due to the poor performance of the chlorophyll-a only models ($R^2$ between -0.3 -- -0.19), resulting in an improvement in $R^2$ of approximately 0.65 -- 0.7. The relative poor performance of zeaxanthin models across all model types was consistent with the weak correlation between zeaxanthin and chlorophyll-\textit{a} (Figure~\ref{fig:pig_corr}) and suggested that this pigment was not strongly expressed in the spectral information captured by the available reflectance bands.

Overall, the pigment models that incorporate spectral information performed best, with the TabPFN model providing modest performance improvements over the RF model across all eight pigments. However, the better performance of the TabPFN model (i.e., higher $R^2$, but similar bias and variance) was relatively small considering the computational cost of this type of model. For the remainder of this analysis, we therefore focus on the RF model that incorporates spectral information.

\subsection{Interpretation of pigment models}
\begin{figure}[htbp]
\centering
\begin{tabular}{cc}
\includegraphics[width=0.44\linewidth]{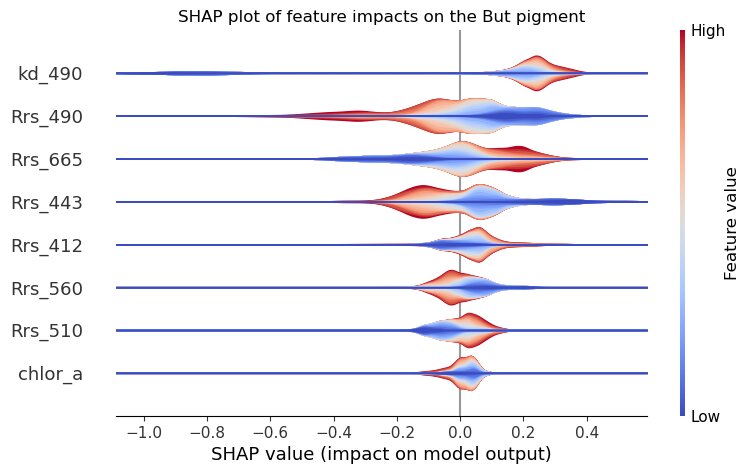} &
\includegraphics[width=0.44\linewidth]{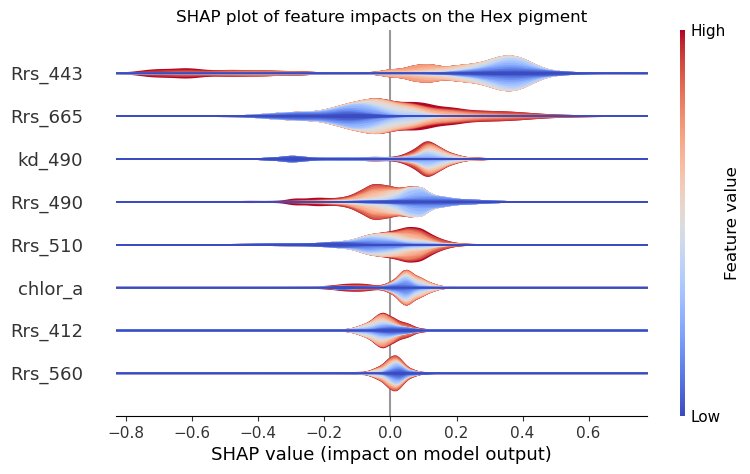} \\
\small (a) 19'-butanoyloxyfucoxanthin & \small (b) 19'-hexanoyloxyfucoxanthin \\

\includegraphics[width=0.44\linewidth]{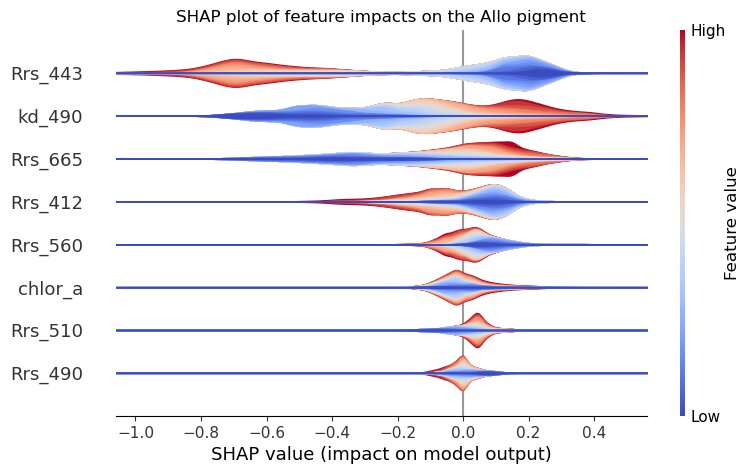} &
\includegraphics[width=0.44\linewidth]{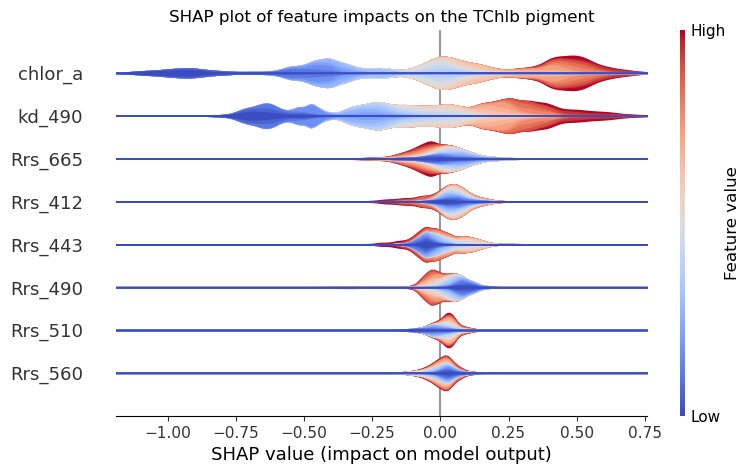} \\
\small (c) Alloxanthin & \small (d) Chlorophyll\textit{-b} \\

\includegraphics[width=0.44\linewidth]{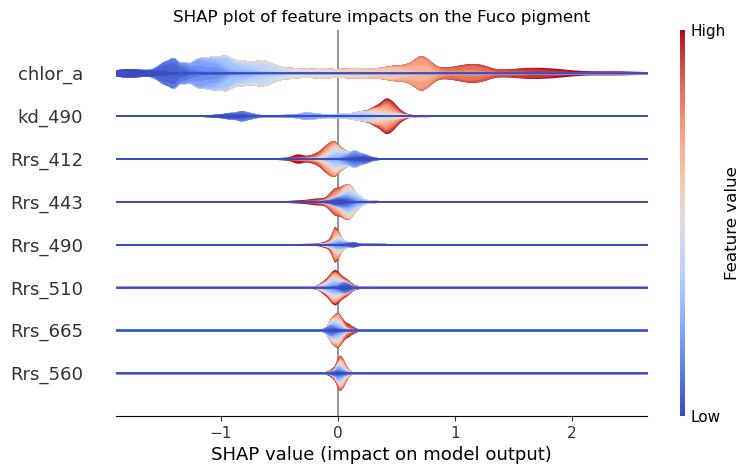} &
\includegraphics[width=0.44\linewidth]{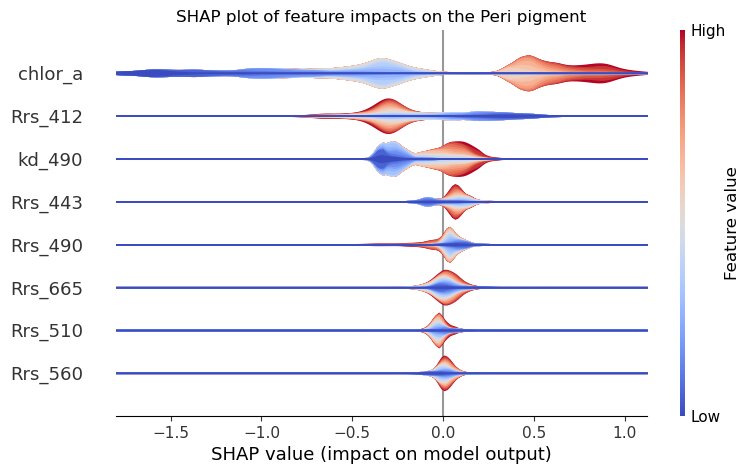} \\
\small (e) Fucoxanthin & \small (f) Peridinin \\

\includegraphics[width=0.44\linewidth]{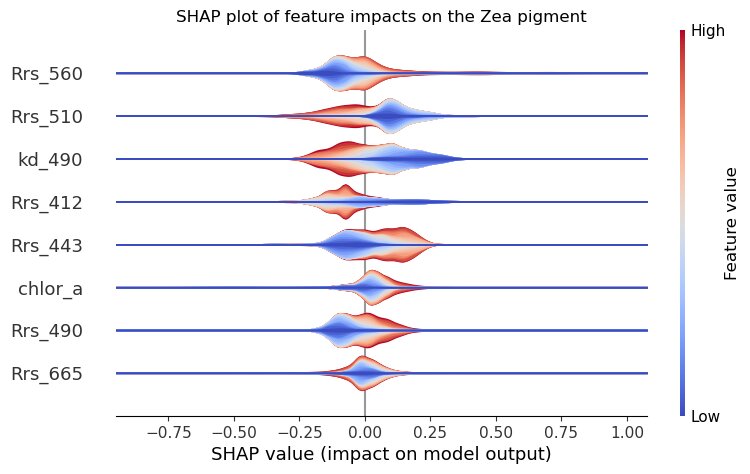} \\
\small (g) Zeaxanthin &  \\
\end{tabular}
\caption{\small SHAP Explainability plots showing the impact of each input variable on pigment concentrations. The x-axis represents the change in predicted pigment concentrations on a logarithmic scale, while the y-axis lists the input features used in the RF models. Features are ordered by importance, with the top feature having the greatest influence. The colours indicate the distribution of feature values and their contribution to the predicted pigment concentrations. Panels (a–g) correspond to the different pigments.}
\label{fig:shap_combined}
\end{figure}

The SHAP analysis revealed substantial differences in the predictor variables utilised by the RF pigment models that incorporated spectral information (Figure~\ref{fig:shap_combined}). Fucoxanthin and peridinin predictions were strongly influenced by chlorophyll-\textit{a}, consistent with the strong correlations observed between these pigments and chlorophyll-\textit{a} in the match-up dataset (Figure~\ref{fig:pig_corr}). For both pigments, chlorophyll-\textit{a} exhibited the largest SHAP values and therefore contributed most to the model predictions (Figure~\ref{fig:shap_combined}). In contrast, the alloxanthin, 19'-butanoyloxyfucoxanthin and 19'-hexanoyloxyfucoxanthin models relied primarily on combinations of $R_{rs}$ bands and $K_{d}(490)$, with chlorophyll-\textit{a} contributing relatively little. The dominance of spectral variables in these models is consistent with the substantial improvement in predictive performance relative to the baseline models that only include chlorophyll\textit{-a}, indicating that additional information relevant to pigment composition is present within the multispectral reflectance observations. 
Zeaxanthin exhibited the most distinct SHAP structure. Despite its comparatively poor predictive performance, the model relied predominantly on blue-green reflectance bands and $K_{d}(490)$ rather than chlorophyll-\textit{a}. As chlorophyll-\textit{a} is typically calculated as a ratio of blue to green reflectance bands, then perhaps the Zeaxanthin pigment is related more to the magnitude of the $R_{rs}$ bands rather than simply the ratio between the bands.
These results suggest that the model is attempting to exploit optical signatures associated with zeaxanthin rather than relying on covariance with chlorophyll\textit{-a}. These results indicated that the machine learning models utilised varying combinations of spectral information depending on the pigment being predicted.

\subsection{Global maps of pigments}

\begin{figure}[htbp] 
\centering 
\footnotesize 
\begin{tabular}{cc} 

\includegraphics[width=\mapwidth,clip,trim=7 6 8 39]{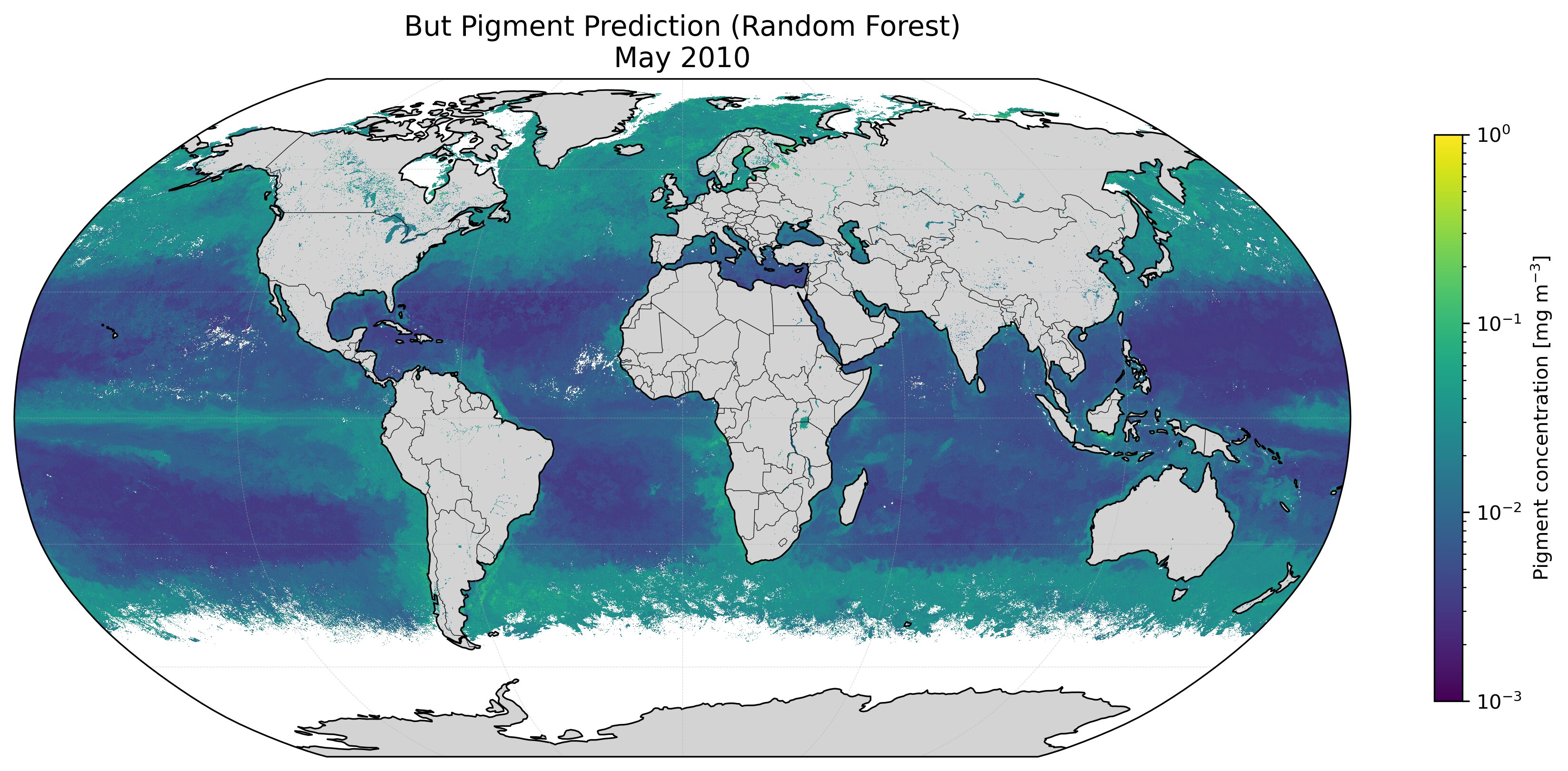} & \includegraphics[width=\mapwidth,clip,trim=7 6 8 39]{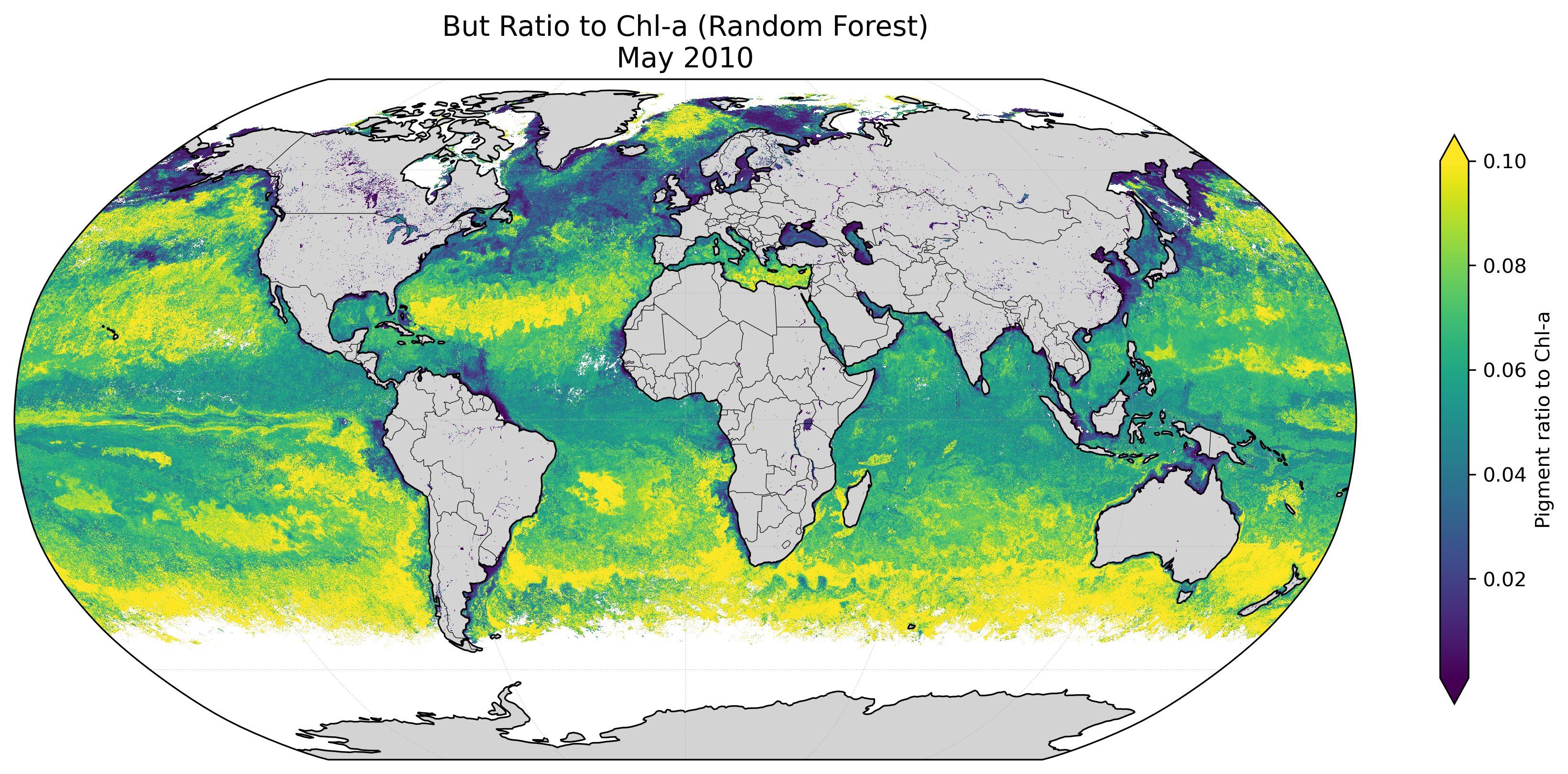} \\ 
\small (a) 19'-butanoyloxyfucoxanthin concentration & \small (b) 19'-butanoyloxyfucoxanthin:Chl-\textit{a} ratio \\ \\

\includegraphics[width=\mapwidth,clip,trim=7 6 8 39]{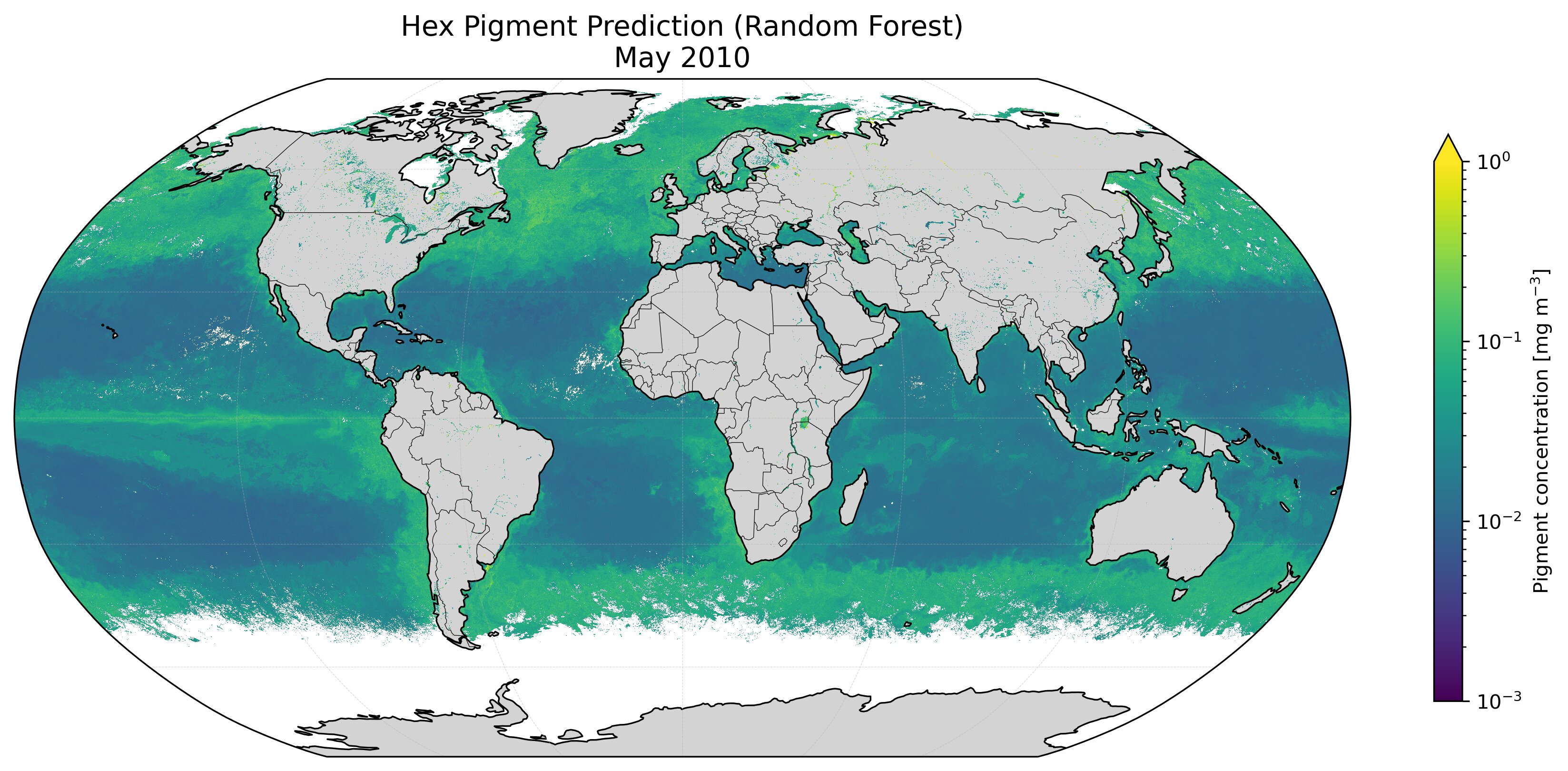} & \includegraphics[width=\mapwidth,clip,trim=7 6 8 39]{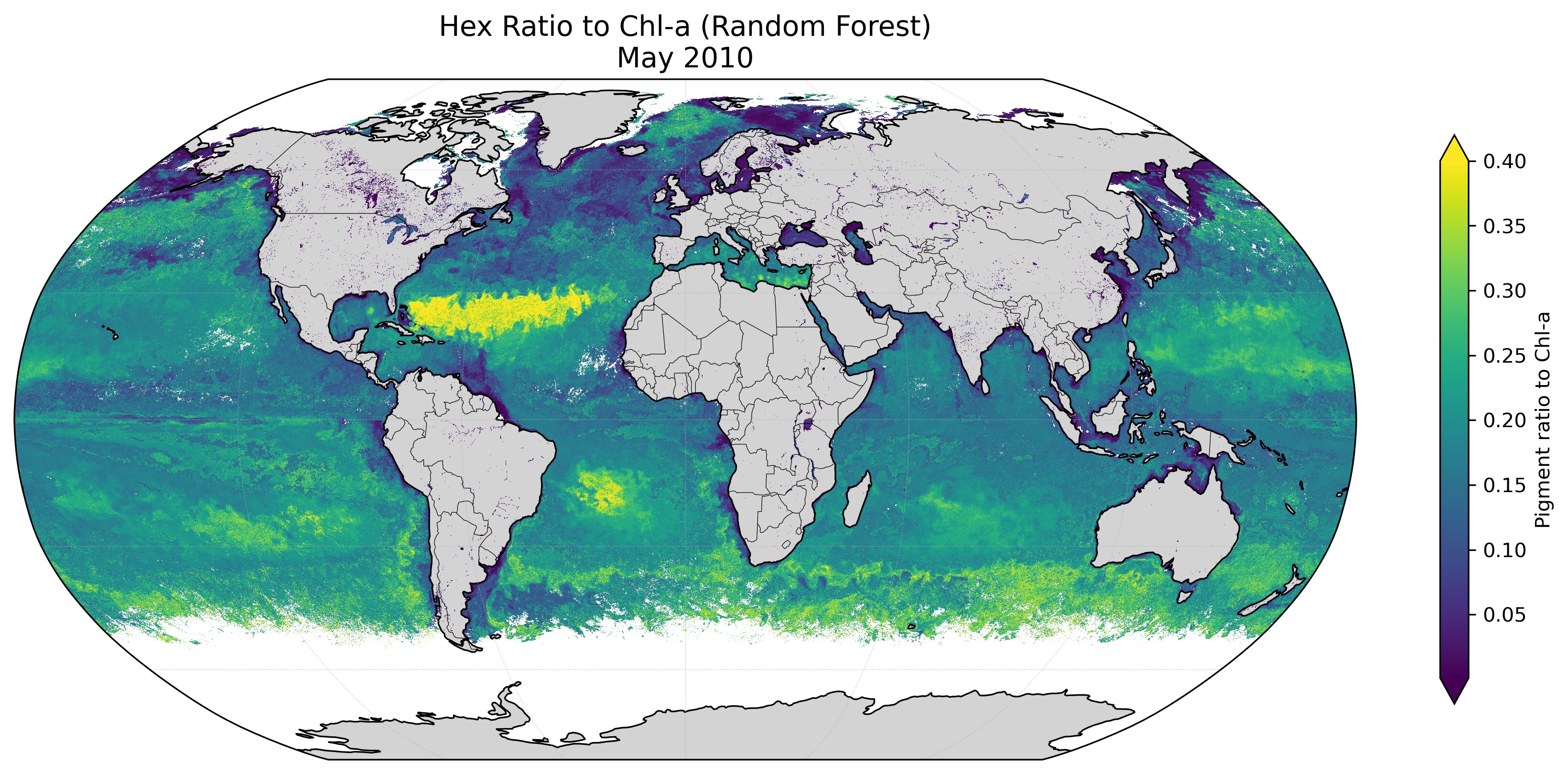} \\
\small (c) 19'-hexanoyloxyfucoxanthin concentration & \small (d) 19'-hexanoyloxyfucoxanthin:Chl-\textit{a} ratio \\ \\ 

\includegraphics[width=\mapwidth,clip,trim=7 6 8 39]{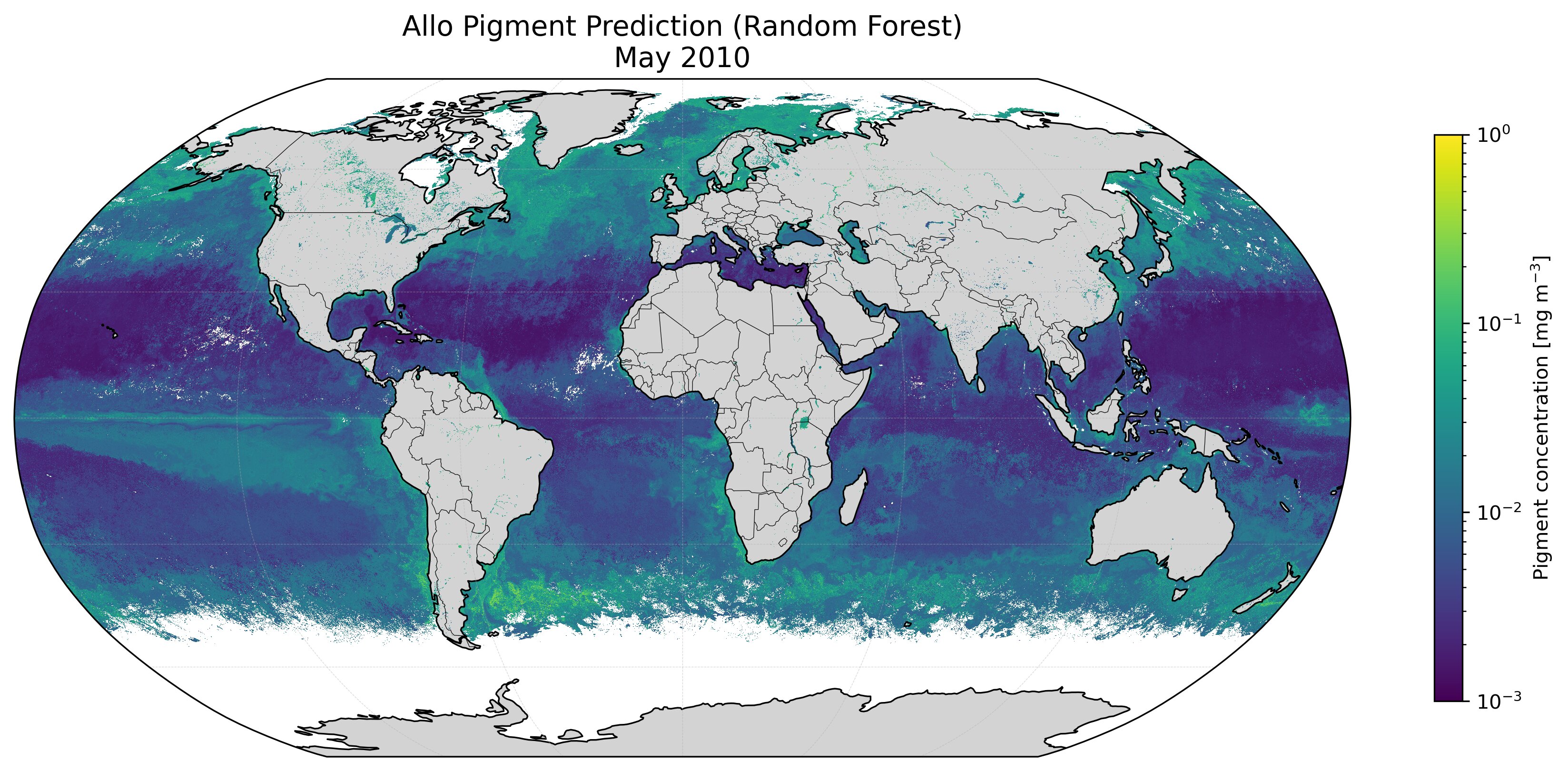} & \includegraphics[width=\mapwidth,clip,trim=7 6 8 39]{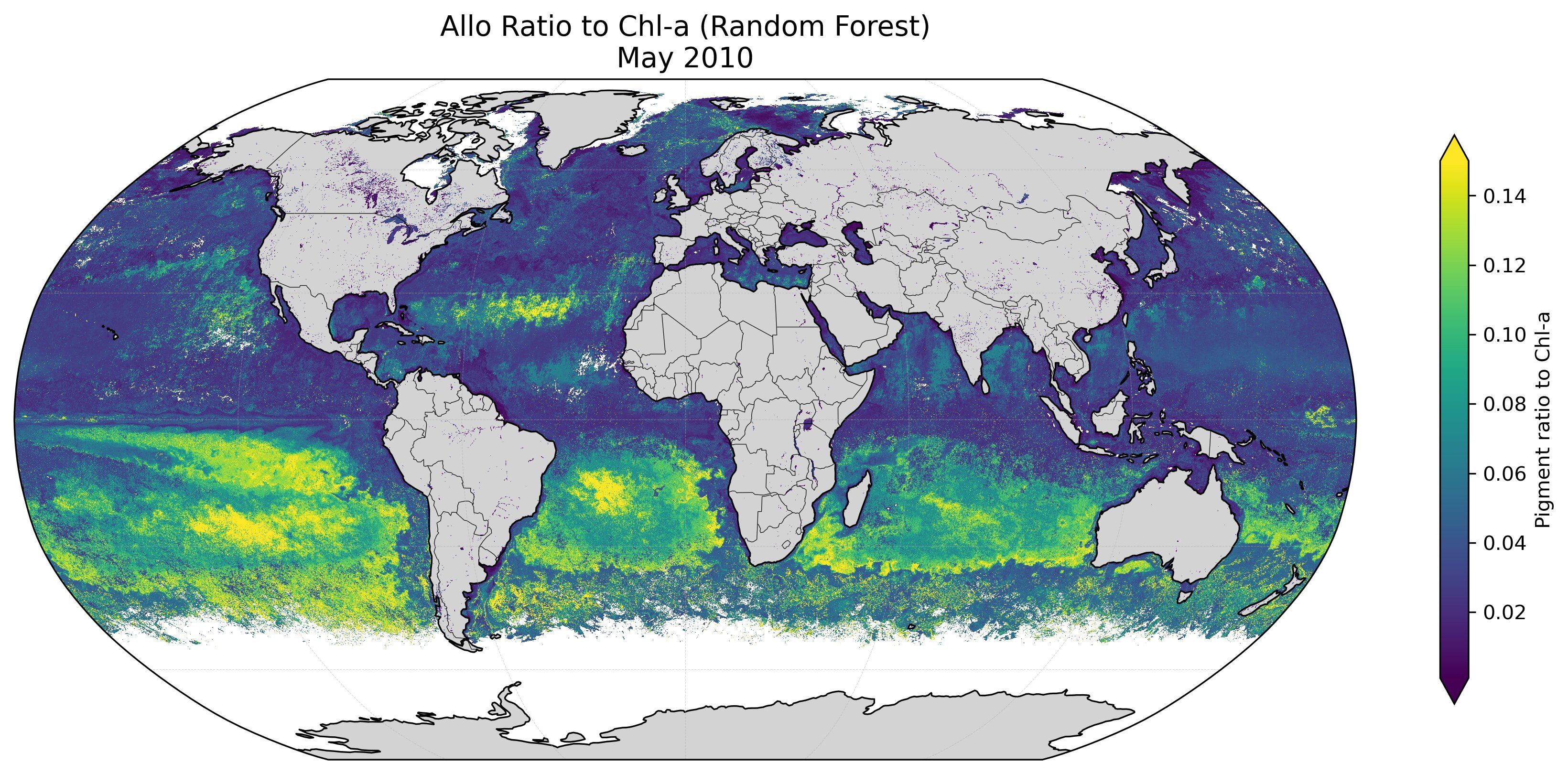} \\
\small (e) Alloxanthin concentration & \small (f) Alloxanthin:Chl-\textit{a} ratio \\ \\ 

\includegraphics[width=\mapwidth,clip,trim=7 6 8 39]{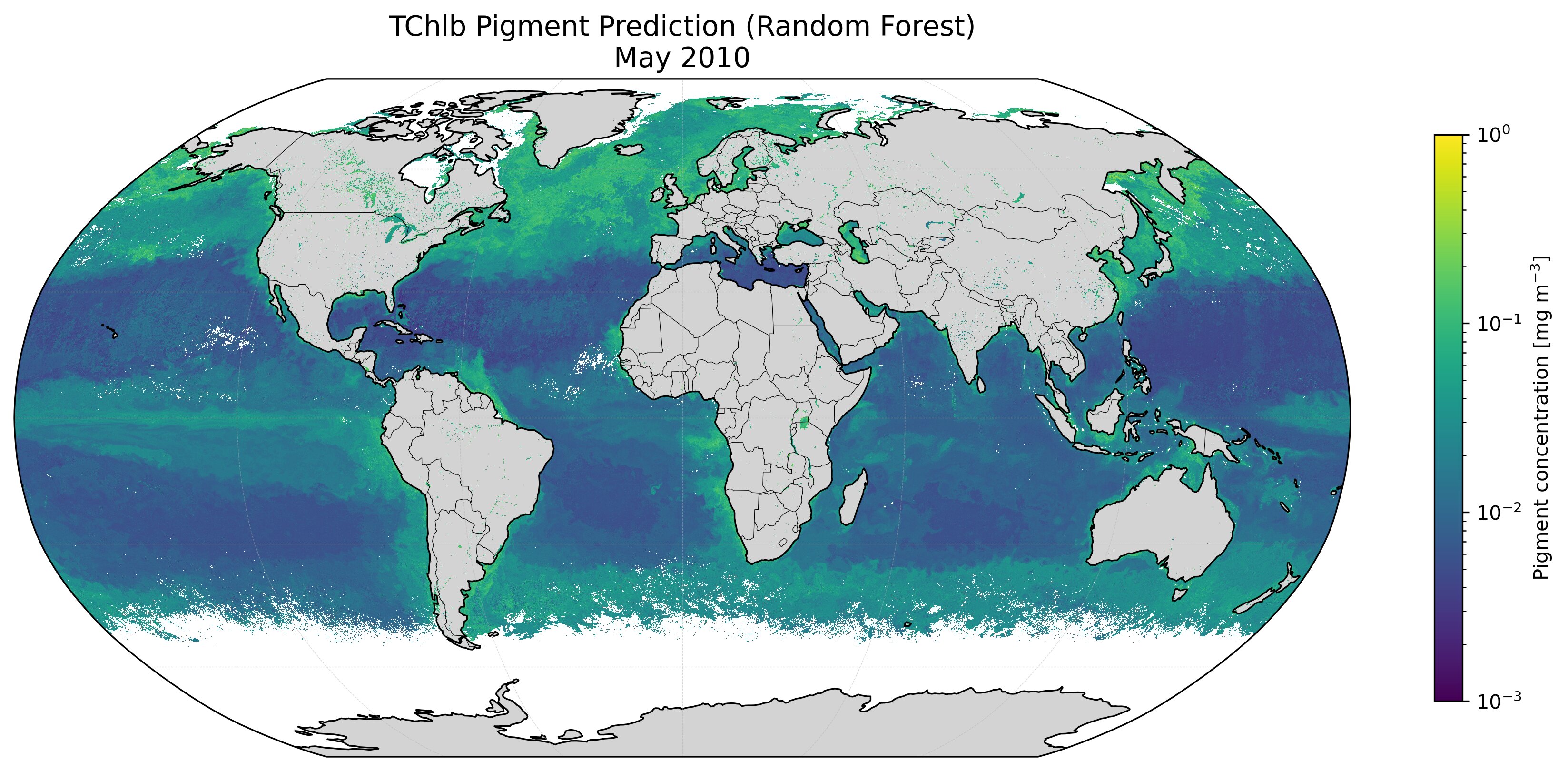} & \includegraphics[width=\mapwidth,clip,trim=7 6 8 39]{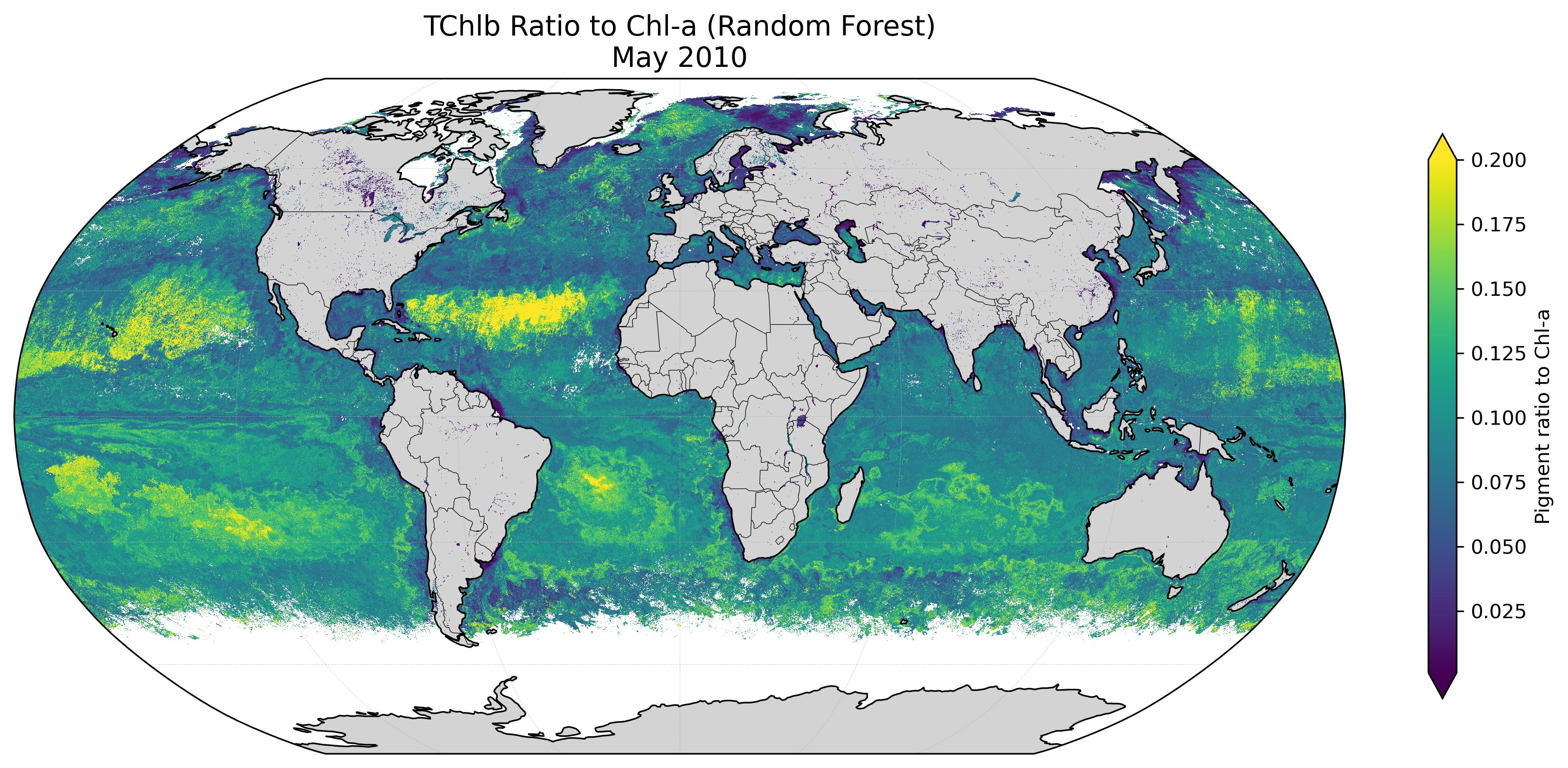} \\ 
\small (g) Total chlorophyll-\textit{b} concentration & \small (h) Total chlorophyll-\textit{b}:Chl-\textit{a} ratio \\

\end{tabular} 
\caption{\small Global maps of predicted pigment concentrations (left column) and pigment-to-chlorophyll-\textit{a} ratios (right column) for May 2010 generated using the Random Forest model that incorporates spectral information. Panels (a–h) show 19'-butanoyloxyfucoxanthin, 19'-hexanoyloxyfucoxanthin, alloxanthin and total chlorophyll-\textit{b}.} 
\label{fig:global_maps_combined_1} 
\end{figure}

\begin{figure}[htbp] 
\centering 
\footnotesize 
\begin{tabular}{cc}

\includegraphics[width=\mapwidth,clip,trim=7 6 8 39]{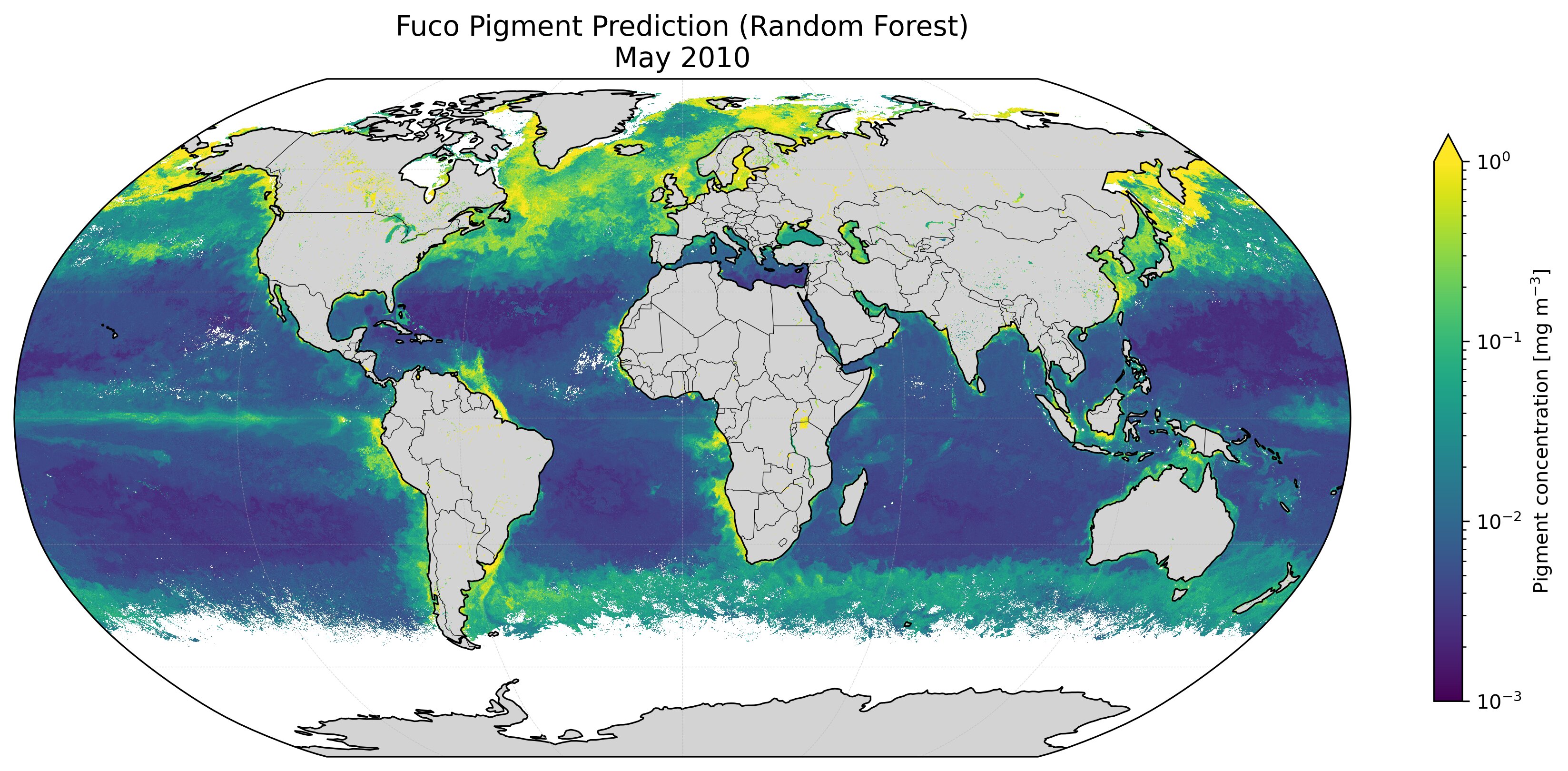} & \includegraphics[width=\mapwidth,clip,trim=7 6 8 39]{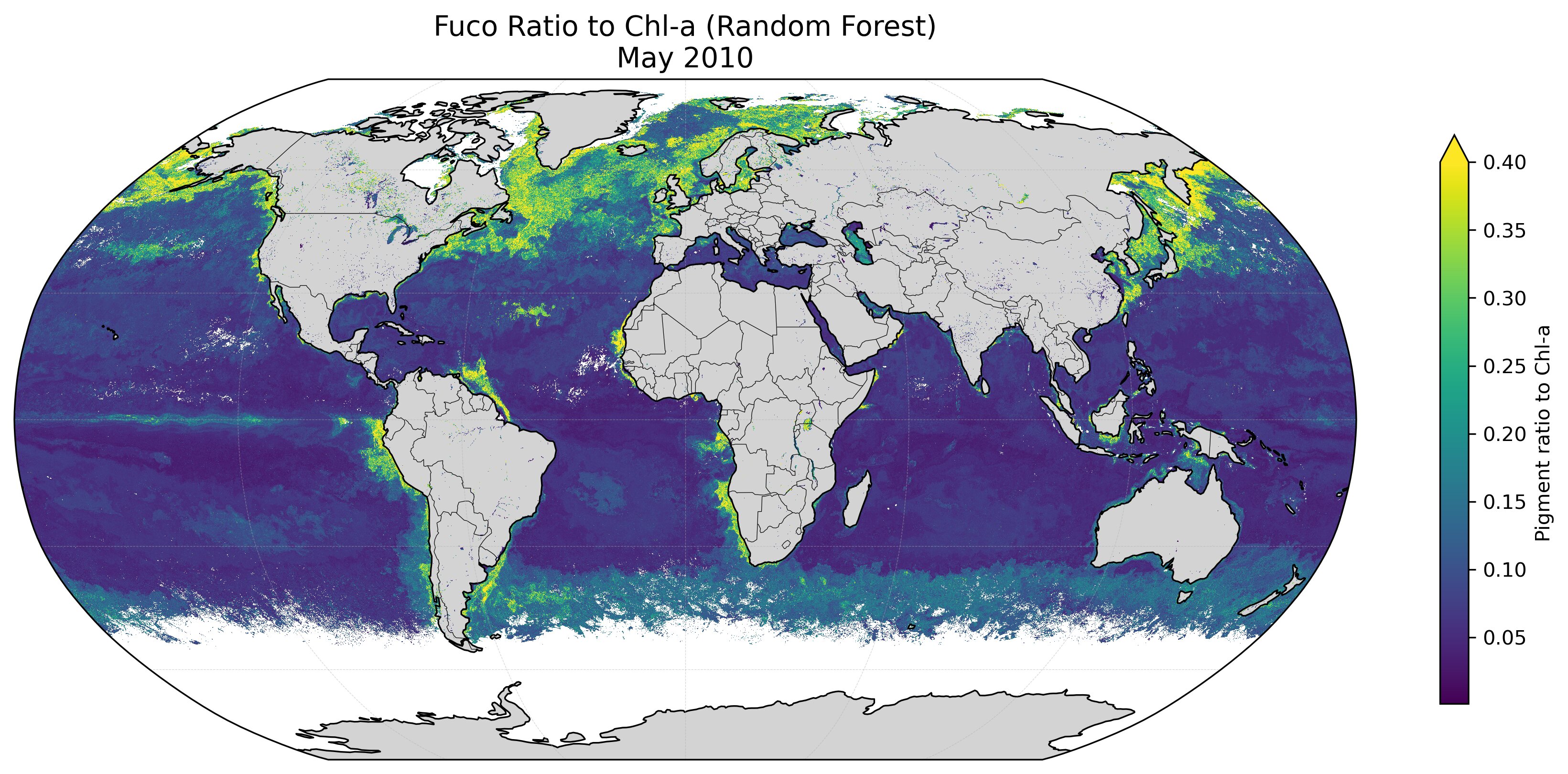} \\ 
\small (a) Fucoxanthin concentration & \small (b) Fucoxanthin:Chl-\textit{a} ratio \\ \\ 

\includegraphics[width=\mapwidth,clip,trim=7 6 8 39]{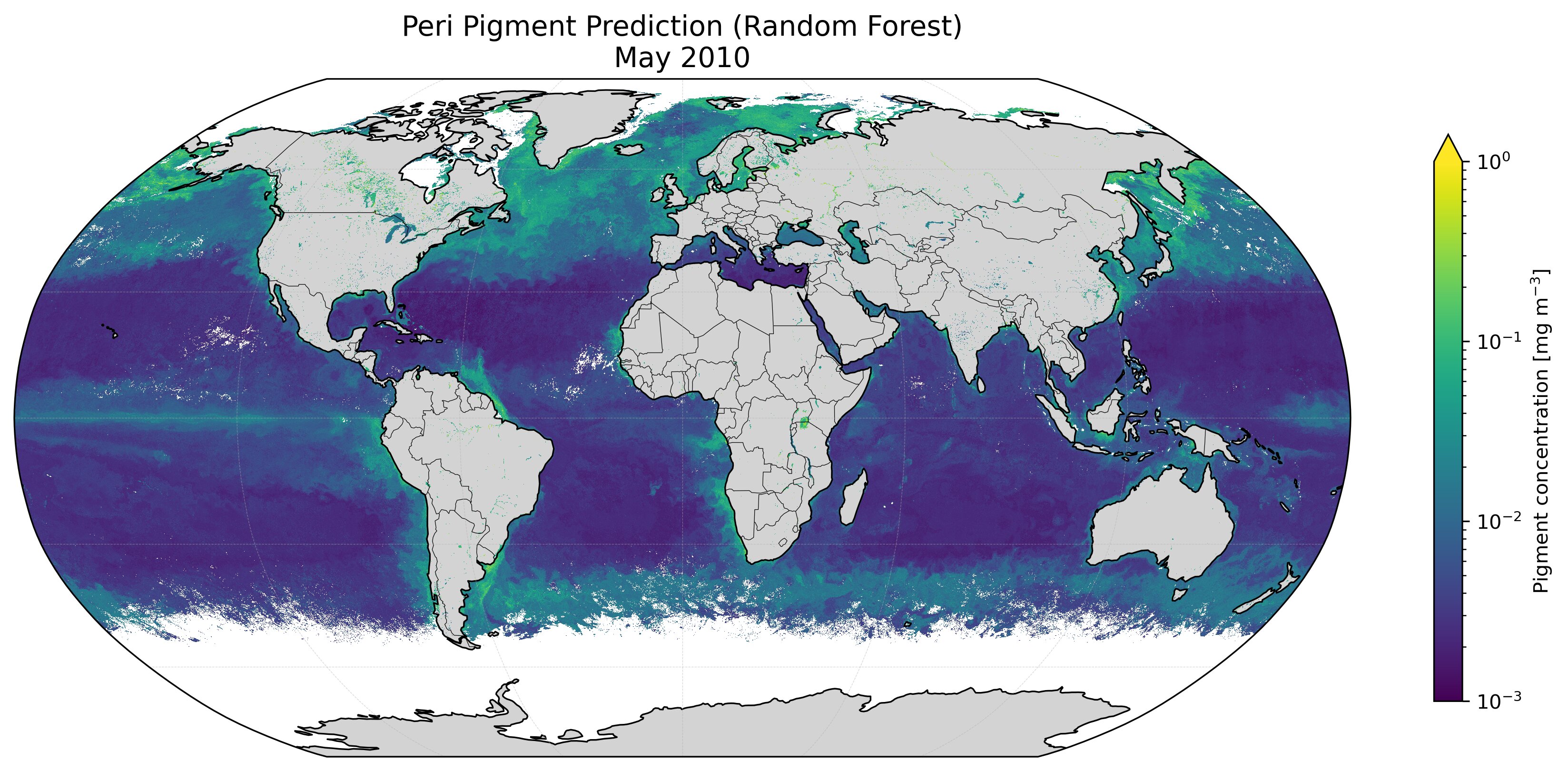} & \includegraphics[width=\mapwidth,clip,trim=7 6 8 39]{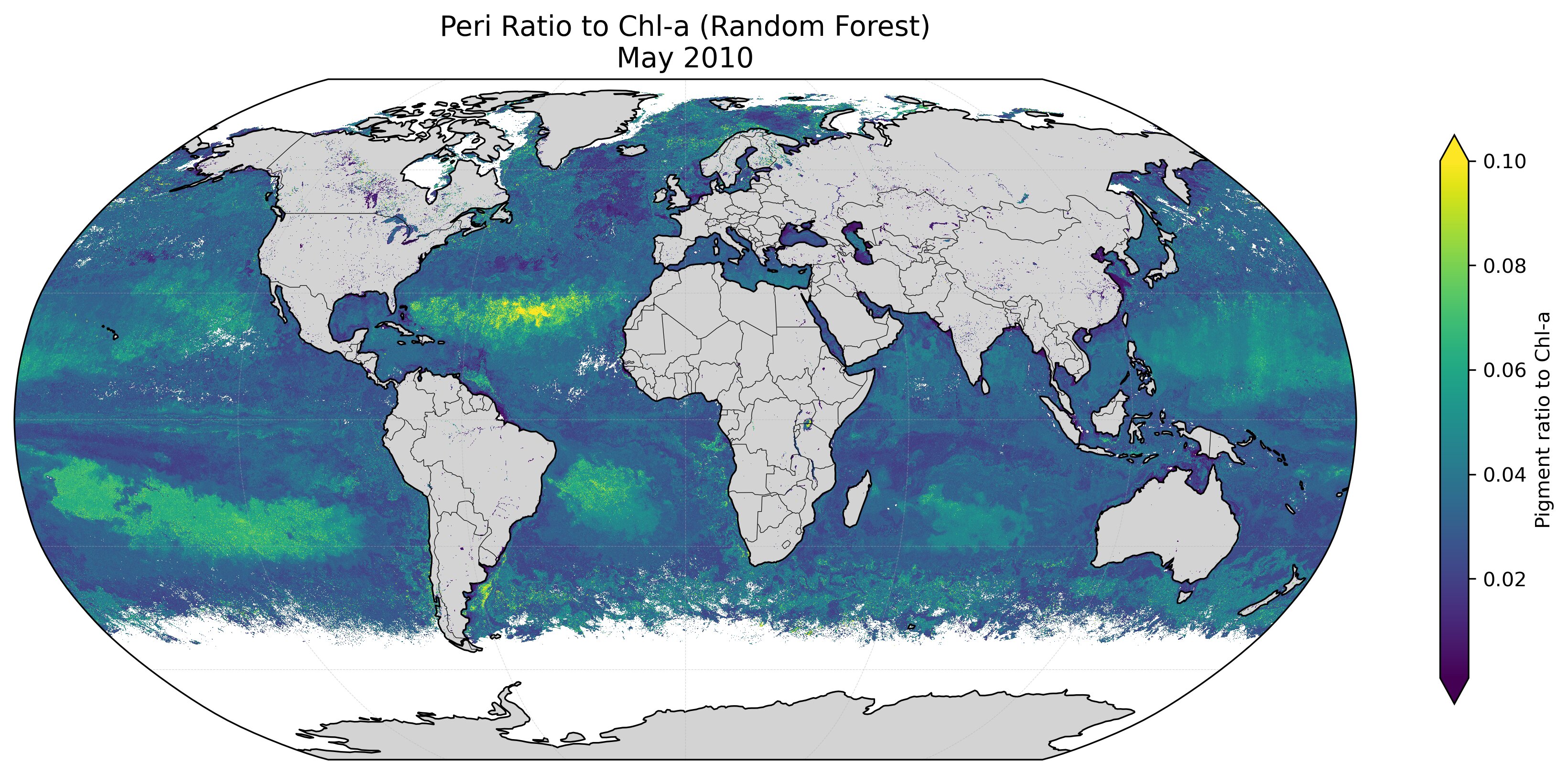} \\ 
\small (c) Peridinin concentration & \small (d) Peridinin:Chl-\textit{a} ratio \\ \\ 

\includegraphics[width=\mapwidth,clip,trim=7 6 8 39]{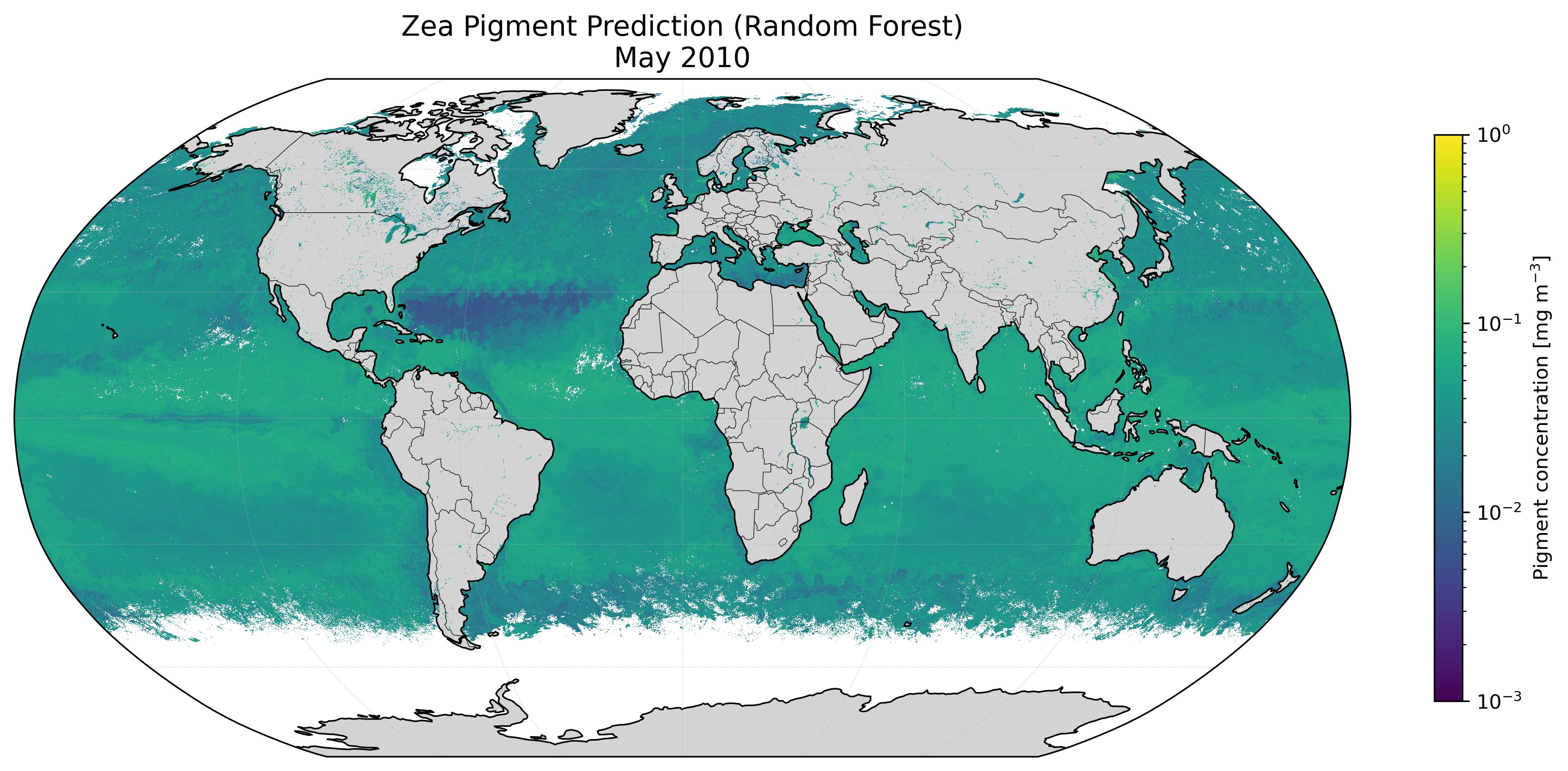} & \includegraphics[width=\mapwidth,clip,trim=7 6 8 39]{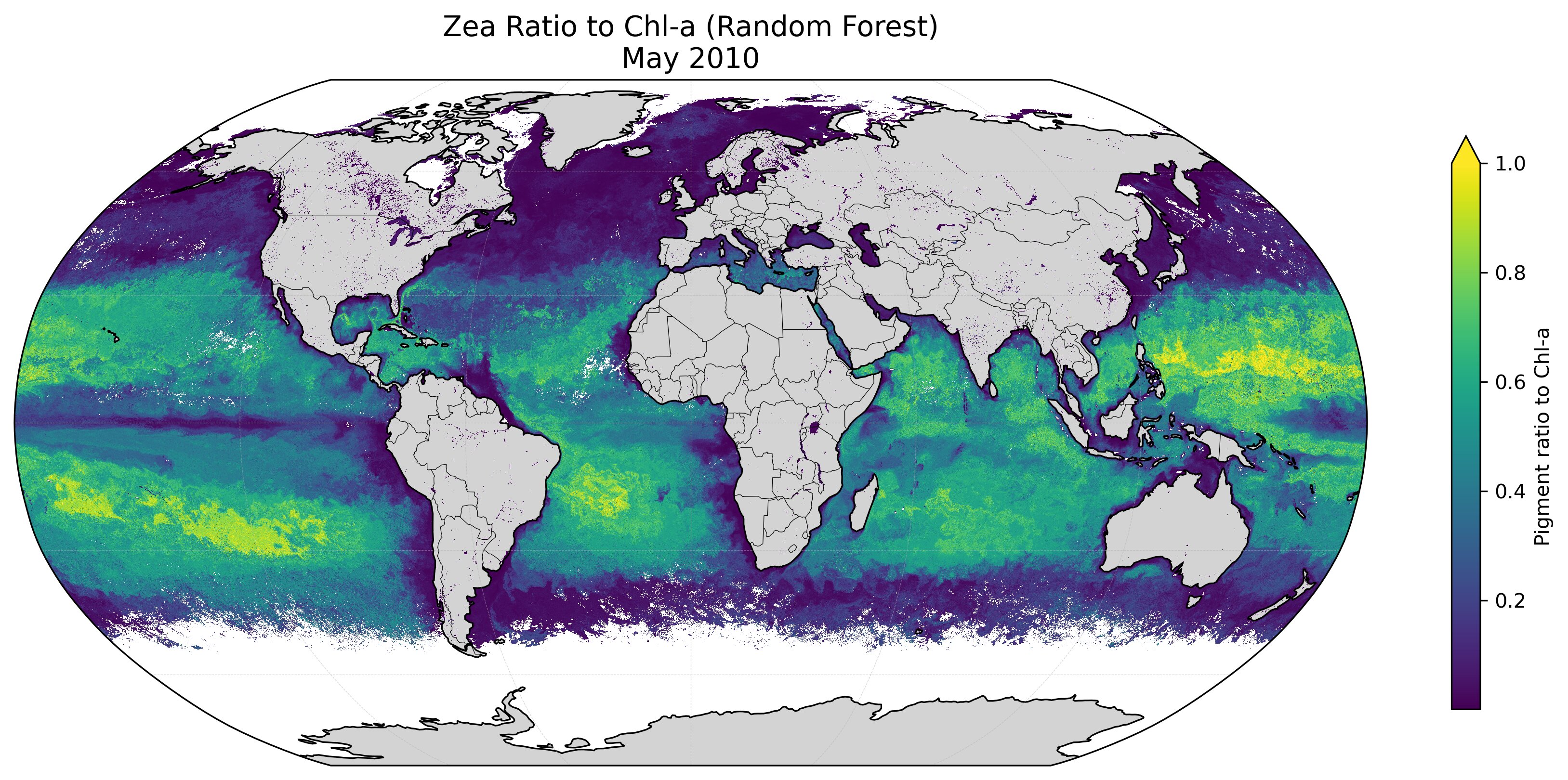} \\ \small (e) Zeaxanthin concentration & \small (f) Zeaxanthin:Chl-\textit{a} ratio \\ \\ 

\end{tabular} 
\caption{\small Global maps of predicted pigment concentrations (left column) and pigment-to-chlorophyll-\textit{a} ratios (right column) for May 2010 generated using the Random Forest model. Panels (a–f) show fucoxanthin, peridinin and zeaxanthin.} 
\label{fig:global_maps_combined_2}
\end{figure}

One of the objectives of this work was to assess whether the machine learning tools tested here could bring in more information about pigment composition than what could be achieved solely on the basis of correlation between chlorophyll-{\it a} and the accessory pigments. Figure~\ref{fig:r2_comparison} provides a quantitative measure of the improvement of the RF model, relative to approaches that rely solely on chlorophyll-{\it a}, when considering the validation data. Additional insights into the algorithm performance can be garnered by studying the global maps of the accessory pigments, in absolute units, and in concentrations relative to chlorophyll-{\it a} (Figure~\ref{fig:global_maps_combined_1} and~\ref{fig:global_maps_combined_2}). The left-hand panels of these two figures show coherent large-scale oceanographic structures, reproducing known contrasts between productive high-latitude and coastal waters and the oligotrophic subtropical gyres. The predicted distributions were therefore consistent with established patterns of marine biogeography and suggest that the machine learning models recovered ecologically meaningful information from multispectral ocean colour observations. While that is reassuring, and is essential for any successful set of algorithms, it is difficult to discern from these panels whether the machine learning models developed here successfully distinguishes each pigment from all the other pigments, and importantly, whether they capture regional differences in the distribution of the accessory pigments. 

This aspect of the investigation was brought to the fore in the right-hand panels of Figure~\ref{fig:global_maps_combined_1} and Figure~\ref{fig:global_maps_combined_2}, in which each accessory pigment is plotted relative to the chlorophyll concentration. These figures clearly demonstrate that the models were able to capture faithfully the shifting community structure of phytoplankton at the global scale, as brought into evidence by the accessory pigments, with, admittedly, some ambiguities when it comes to relating pigments to phytoplankton types. To highlight some notable features, the relative concentration of fucoxanthin has a global distribution that mirrors that of the absolute concentration of fucoxanthin (Figure~\ref{fig:global_maps_combined_2}a and b), consistent with the increase in diatoms with increasing chlorophyll-{\it a} concentrations. But this result is not a true test of independence from chlorophyll-{\it a} variability, since we know that fucoxanthin is highly correlated with chlorophyll-{\it a} (Figure~\ref{fig:pig_corr}). From this perspective, it is more interesting to see the relative distribution of total chlorophyll-{\it b} (Figure ~\ref{fig:global_maps_combined_1}h). As expected based on the known distribution of {\it Prochlorococcus}, whose concentrations are highest in the subtropical gyres, and whose pigment complement includes divinyl chlorophyll-{\it b}, the results captured this feature very well. The pigment dataset that we are using combines divinyl chlorophyll-{\it b} and regular chlorophyll-{\it b} into a total chlorophyll-{\it b}, such that some of the chlorophyll-{\it b} we see outside of the subtropical gyres may be associated with chlorophytes rather than \textit{Prochlorococcus}. Similarly, relative concentrations of 19'-butanoyloxyfucocanthin (Figure~\ref{fig:global_maps_combined_1}b) considered to be marker pigments of haptophytes, revealed their predominance in the Southern Hemisphere, where as and 19-hexanoyloxyfucoxanthin(Figure~\ref{fig:global_maps_combined_1}d), also considered to be marker pigments of haptophytes tends to be present across both hemispheres, which is supported by existing literature~\citep{liu2009extreme}. 
The performance of the zeaxanthin retrieval was relatively poor when compared with the test data (Figure~\ref{fig:result_scatter_2}d), which showed that the machine learning models do not capture much more than the mean of its concentration. This quality was reflected in Figure~\ref{fig:global_maps_combined_1}e, where zeaxanthin concentration appeared to be relatively uniformly distributed over the whole globe. Nevertheless, the distribution of the relative concentration of zeaxanthin (Figure~\ref{fig:global_maps_combined_2}f) revealed a more realistic pattern. With the exception of zeaxanthin, the predicted distributions were therefore consistent with established patterns of marine biogeography and suggest that the machine learning models recovered ecologically meaningful information from multispectral ocean colour observations. 

Several pigments exhibited spatial patterns that closely resembled those of total chlorophyll-\textit{a}, which is shown in the chlorophyll-\textit{a} to pigment ratio plots (Figure~\ref{fig:global_maps_combined_1}b,d,f,h and Figure~\ref{fig:global_maps_combined_1}b,d,f). Fucoxanthin and peridinin showed strong enhancement in productive regions, including the North Atlantic, North Pacific, Southern Ocean and major coastal upwelling systems, while concentrations were substantially reduced within the subtropical gyres. These distributions are consistent with the ecology of larger phytoplankton groups, including diatoms and dinoflagellates, that characteristically dominate nutrient-rich waters. The similarity between the fucoxanthin, peridinin and chlorophyll-\textit{a} maps is also consistent with the strong chlorophyll dependence identified by the SHAP analysis and the comparatively limited improvement over chlorophyll-only baseline models. 
A second group of pigments, including alloxanthin, 19'-butanoyloxyfucoxanthin and total chlorophyll-\textit{b}, displayed intermediate behaviour. These pigments remained associated with productive waters but exhibited weaker large-scale gradients and increased regional variability relative to fucoxanthin and peridinin. Elevated concentrations were observed at high latitudes and along equatorial regions, whereas reduced concentrations persisted in the subtropical gyres. These pigments are commonly associated with cryptophytes, chlorophytes, prasinophytes, haptophytes and pelagophytes, suggesting that the models are capturing variability associated with phytoplankton groups occupying a broader range of ecological niches.
In contrast, 19'-hexanoyloxyfucoxanthin and zeaxanthin exhibited substantially broader open-ocean distributions. For 19'-hexanoyloxyfucoxanthin, elevated concentrations remained evident at high latitudes but the reduction in oligotrophic gyres was less pronounced than the other accessory pigments. The resulting distribution was more spatially continuous across ocean basins, consistent with the widespread occurrence of haptophyte communities throughout the global ocean~\citep{liu2009extreme}. Zeaxanthin displayed the most distinctive pattern of all pigments, with comparatively weak large-scale gradients and relatively high concentrations persisting throughout tropical and subtropical waters. This pattern contrasts strongly with that of chlorophyll-\textit{a} and is consistent with the global distribution of picophytoplankton communities, including \textit{Prochlorococcus} and \textit{Synechococcus}, which are adapted to stratified oligotrophic environments.

In comparison with total chlorophyll-\textit{a}, if pigment concentrations were being estimated solely from their covariance with chlorophyll-\textit{a}, all maps would be expected to exhibit similar spatial patterns. Instead, substantial differences are observed among pigments, particularly between fucoxanthin, peridinin and chlorophyll-\textit{a} on one hand, and 19'-hexanoyloxyfucoxanthin and zeaxanthin on the other. These contrasting distributions suggest that the models are exploiting information beyond chlorophyll-\textit{a} and are capable of recovering broad-scale variability in phytoplankton community composition from multispectral ocean colour observations.

\section{Discussion} 
The primary aim of this study was to evaluate whether multispectral ocean colour observations contain information that can improve estimates of diagnostic pigments beyond that provided by chlorophyll-\textit{a} alone. Across all pigments examined, machine learning models trained using multispectral reflectance consistently outperformed models trained solely on chlorophyll-\textit{a}. This demonstrates that ocean colour reflectance contains additional information relevant to phytoplankton pigment composition, although the magnitude of this information varies substantially between pigments.

The strongest predictive performance was obtained for fucoxanthin. However, fucoxanthin also exhibited the smallest improvement relative to the baseline model, which only had chlorophyll-a as an input. This result is consistent with both the correlation analysis and the SHAP interpretation. Fucoxanthin showed the strongest correlation with chlorophyll-\textit{a} ($r=0.95$), and the SHAP analysis revealed that chlorophyll-\textit{a} dominates model predictions. Taken together, these results suggest that a large proportion of fucoxanthin predictability arises from its covariance with total chlorophyll-\textit{a} concentration.

In contrast, models of alloxanthin, 19'-butanoyloxyfucoxanthin and 19'-hexanoyloxyfucoxanthin that incorporated spectral information showed substantially larger improvements relative to the baseline models. For these pigments, the SHAP analysis demonstrated that multispectral reflectance bands and $K_{d}(490)$ were generally more influential than chlorophyll-\textit{a}. The consistency between the SHAP rankings and model performance provides strong evidence that predictive skill for these pigments derives from information contained within the spectral reflectance observations rather than from chlorophyll-\textit{a} alone. This supports the hypothesis that multispectral ocean colour observations contain information related to phytoplankton community composition that is not fully represented by chlorophyll-\textit{a}.

The results for zeaxanthin are particularly informative. Zeaxanthin was the least accurately predicted pigment and exhibited only weak correlation with chlorophyll-\textit{a}. Nevertheless, the SHAP analysis revealed that model predictions depend primarily on blue-green reflectance bands and $K_{d}(490)$ rather than chlorophyll-\textit{a}. Furthermore, the global map of zeaxanthin differed markedly from those of other pigments, exhibiting a broader distribution across tropical and subtropical waters as might be expected for smaller phytoplankton species that contain this pigment. Although the overall predictive accuracy remained modest, these results suggest that the model is identifying an ecological signal distinct from total phytoplankton biomass. This finding is encouraging because it implies that ocean colour observations may contain information about cyanobacteria-associated pigments that is not captured by chlorophyll-a-based approaches. It has previously been shown that satellite based sea surface temperature can be used to improve the prediction of Zea pigment~\citep{pan2013remote}. This could be considered for future work.

The global maps provided an additional line of evidence supporting this interpretation. The predicted distributions display coherent large scale biogeographical patterns that are consistent with known contrasts between productive high-latitude waters and oligotrophic subtropical gyres. Pigments such as fucoxanthin, peridinin and alloxanthin are typically found in larger phytoplankton cells, including diatoms, dinoflagellates and haptophytes, which have a higher nutrient demand and are therefore found in regions where nutrient concentrations are higher, i.e., coastal, upwelling and seasonally stratified regions~\citep{kramer_how_2019}. Pigments such as zeaxanthin are typically found in smaller phytoplankton cells, including \textit{Prochlorococcus}, that can benefit from relatively low nutrient concentrations in open oceans. The spatial distribution of the pigments is therefore consistent with contrasting ecological niches occupied by the underlying phytoplankton communities. Importantly, the maps differ substantially between pigments. If the models were simply reproducing chlorophyll-\textit{a} variability and scaling pigments according to their covariance with biomass, more similar spatial patterns would be expected. Instead, the observed differences suggest that the machine learning models are extracting information related to phytoplankton community structure and ecological niche variability.

Our results are broadly consistent with previous studies that have demonstrated limited but measurable skill in predicting pigments from remotely sensed observations \citep{bracher_using_2015,el2019estimation,stock_accuracy_2020}. Earlier work has often reported stronger predictive performance than observed here, particularly when random train and test splitting strategies were employed. However, oceanographic datasets exhibit substantial spatial and temporal autocorrelation, meaning that nearby samples are frequently more similar than independent observations. Consequently, random splitting can lead to information leakage between training and validation datasets and produce overly optimistic performance estimates~\citep{stock_accuracy_2020}. By withholding entire years from model training, our validation strategy provides a more realistic assessment of model generalisation to unseen observations. The reduced performance reported here may therefore better represent the predictive capability achievable in operational applications.

The SHAP analysis further demonstrates the value of explainable AI techniques for ocean colour applications. Rather than treating machine learning models as black boxes, SHAP provides insight into the physical variables driving model predictions and allows comparison between pigments. The distinction between chlorophyll-a dominated pigments, such as fucoxanthin and peridinin, and the pigments where the $R_{rs}$ bands are considered more important, such as alloxanthin, 19'-butanoyloxyfucoxanthin and 19'-hexanoyloxyfucoxanthin, is particularly informative. These results suggest that spectral-anomalies, around the chlorophyll-\textit{a} to $R_{rs}$ relationships, are better associated with certain pigments~\citep{alvain2005remote}

Overall, the results indicate that machine learning methods can extract ecologically meaningful information on phytoplankton pigment composition from multispectral ocean colour observations. While chlorophyll-\textit{a} remains a dominant predictor for some pigments, the inclusion of spectral reflectance data provides measurable improvements in predictive skill and yields physically plausible global distributions. These findings support the use of machine learning approaches as a complementary tool for deriving information on phytoplankton community composition from long-term ocean colour records and provide a foundation for future exploitation of hyperspectral satellite observations.

Our work has only been possible through the collection of in situ HPLC pigment measurements. The algorithms developed here are based on relationships between ocean colour and pigment data collected over past decades, which may change under future ocean conditions~\citep{sathyendranath2017ocean}. Continued collection of these in situ measurements is therefore essential to improve algorithm performance and detect climate-driven changes in model parameterisations~\citep{satterthwaite2025essential}.


\section{Conclusions}
This study evaluated the ability of machine learning models to estimate diagnostic phytoplankton pigments from multispectral ocean colour observations. Models incorporating multispectral reflectance consistently outperformed chlorophyll-only approaches, demonstrating that ocean colour observations contain information on phytoplankton pigment composition beyond that provided by chlorophyll-\textit{a} alone.

Predictive skill varied considerably between pigments. Fucoxanthin achieved the highest overall accuracy but was also the pigment most strongly associated with chlorophyll-\textit{a}, indicating that much of its predictability derives from covariance with phytoplankton biomass. In contrast, alloxanthin, but-fucoxanthin and hex-butfucoxanthin exhibited substantial improvements relative to chlorophyll only models, demonstrating that additional information contained within multispectral reflectance observations can be exploited to improve pigment estimation. Although zeaxanthin remained challenging to predict, its distinct SHAP signatures and global distribution suggest that the model captures ecologically meaningful variability.

The SHAP analysis and global pigment maps provided complementary evidence that the models do not simply reproduce chlorophyll-\textit{a} variability. Instead, different pigments were associated with distinct combinations of spectral and optical predictors, resulting in spatial distributions that are consistent with known patterns of phytoplankton biogeography.

Overall, these results demonstrate that multispectral ocean colour observations contain measurable information on diagnostic pigments beyond chlorophyll-\textit{a}. Machine learning approaches provide a practical framework for exploiting this information and offer a pathway towards improved satellite monitoring of phytoplankton community composition in the era of hyperspectral Earth Observation.

%
%
%

\section*{Funding}
This research was supported by the European Space Agency (ESA) through the Biodiversity in the Open Ocean: Mapping, Monitoring and Modelling (BOOMS, grant number 4000137125/22/I-DT ) project and the Biodiversity of the Coastal Ocean: Monitoring with Earth Observation (BiCOME, grant number 4000135756/21/I-EF ) project. Additional support was provided by the Simons Foundation through the Computational Biogeochemical Modeling of Marine Ecosystems (CBIOMES; grant 549947 to S.S.) project and the ESA PHYTOplankton biomass and diversity Climate Change Initiative (PHYTO-CCI; contract number 4000147645/25/I-LR). XS and RB were supported by a UKRI FLF grant (MR/V022792/1). This work is a contribution to the activities of the National Centre of Earth Observation (NCEO) of the UK. 

%
%

\bibliographystyle{Frontiers-Harvard}
\bibliography{pigment-bib}

\end{document}